\documentclass{article}

\usepackage{epsfig}
\usepackage{graphics}
\usepackage{graphicx}
\usepackage[centertags]{amsmath}
\usepackage{amsfonts}
\usepackage{amsthm}
\usepackage{amssymb}
\usepackage{url}
\usepackage{subfigure}
\usepackage{hyperref}

\newenvironment{myitemize}{
\begin{itemize}
  \setlength{\itemsep}{1pt}
  \setlength{\parskip}{0pt}
  \setlength{\parsep}{0pt}}{\end{itemize}
}

\begin{document}

\title{Computation and Universality: Class IV versus Class III Cellular Automata}

\author{Genaro J. Mart{\'i}nez$^{1,2}$, Juan C. Seck-Tuoh-Mora$^3$, and Hector Zenil$^4$}
\date{\it Accepted in August 26, 2012 \\ Published in March 2013 \\ Journal of Cellular Automata, 7(5-6), 393-430, 2013. \\ \url{http://www.oldcitypublishing.com/JCA/JCAcontents/JCAv7n5-6contents.html}}

\maketitle

\begin{centering}
$^1$ Unconventional Computing Center, Department of Computer Science, \\ University of the West of England, Bristol, United Kingdom. \\
\url{genaro.martinez@uwe.ac.uk}\\
$^2$ Departamento de Ciencias e Ingenier{\'i}a de la Computaci{\'o}n, \\ Escuela Superior de C\'omputo, Instituto Polit\'ecnico Nacional, M\'exico. \\
$^3$ \'Area Acad\'emica de Ingenier{\'i}a, \\ Universidad Aut\'onoma del Estado de Hidalgo, M\'exico.\\
\url{jseck@uaeh.edu.mx} \\
$^4$ Department of Computer Science/Kroto Research Institute,\\University of Sheffield, United Kingdom.\\
\url{h.zenil@sheffield.ac.uk} \\
\end{centering}

\begin{abstract}
This paper examines the claim that cellular automata (CA) belonging to {\it Class III} (in Wolfram's classification) are capable of (Turing universal) computation. We explore some chaotic CA (believed to belong to {\it Class III}) reported over the course of the CA history, that may be candidates for universal computation, hence spurring the discussion on Turing universality on both Wolfram's classes III and IV.\\

\end{abstract}

\section{Cellular Automata and Wolfram's classes}
The classification and identification of cellular automata (CA) has become a central focus of research in the field. In \cite{kn:Wolf84}, Stephen Wolfram presented his now well-known {\it classes}. Wolfram's analysis included a thorough study of one-dimensional (1D) CA, order $(k=2,r=2)$ (where $k \in \mathcal{Z^+}$ is the cardinality of the finite alphabet and $r \in \mathcal{Z^+}$ the number of neighbours), and also found the same classes of behaviour in other CA rule spaces. This allowed Wolfram to generalise his classification to all sorts of systems in \cite{kn:Wolf02}. 

An Elementary Cellular Automaton (ECA) is a finite automaton defined in a 1D array. The automaton assumes two states, and updates its state in discrete time according to its own state and the state of its two closest neighbours, all cells updating their states synchronously. \\

Wolfram's classes can be characterised as follows:

\begin{myitemize}
\item {Class I.} CA evolving to a homogeneous state
\item {Class II.} CA evolving periodically
\item {Class III.} CA evolving chaotically
\item {Class IV.} Includes all previous cases, known as a class of {\it complex rules}
\end{myitemize}

Otherwise explained, in the case of a given CA,:

\begin{myitemize}
\item If the evolution is dominated by a unique state of its alphabet for any random initial condition, then it belongs to {\it Class I}.
\item If the evolution is dominated by blocks of cells which are periodically repeated for any random initial condition, then it belongs to {\it Class II}.
\item If for a long time and for any random initial condition, the evolution is dominated by sets of cells without any defined pattern, then it belongs to {\it Class III}.
\item If the evolution is dominated by non-trivial structures emerging and travelling along the evolution space where uniform, periodic, or chaotic regions can coexist with these structures, then it belongs to {\it Class IV}. This class is frequently tagged: {\it complex behaviour}, {\it complexity dynamics}, or simply {\it complex}.
\end{myitemize} 

\begin{figure}
\begin{center}
\subfigure[]{\scalebox{0.47}{\includegraphics{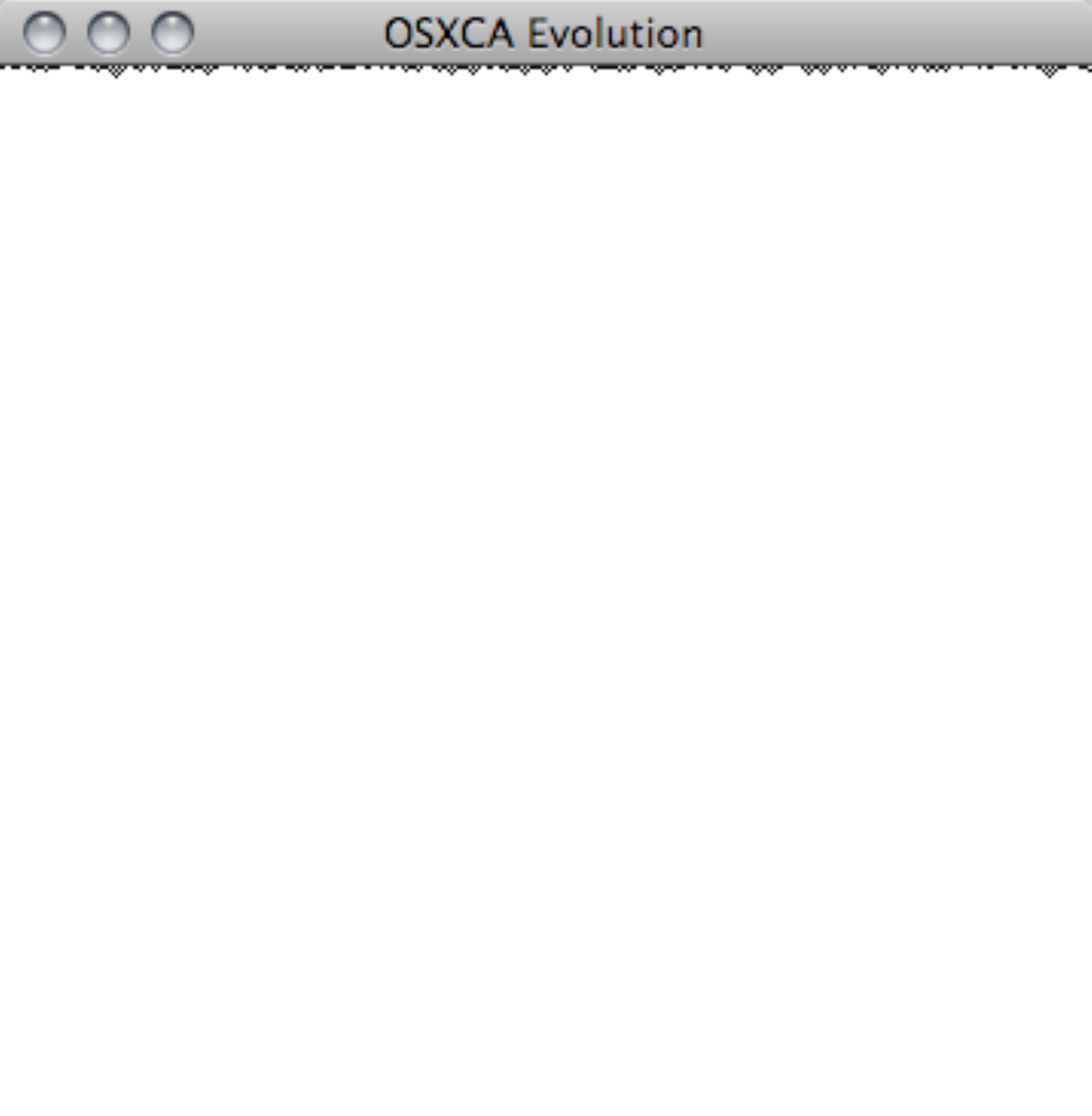}}} 
\subfigure[]{\scalebox{0.47}{\includegraphics{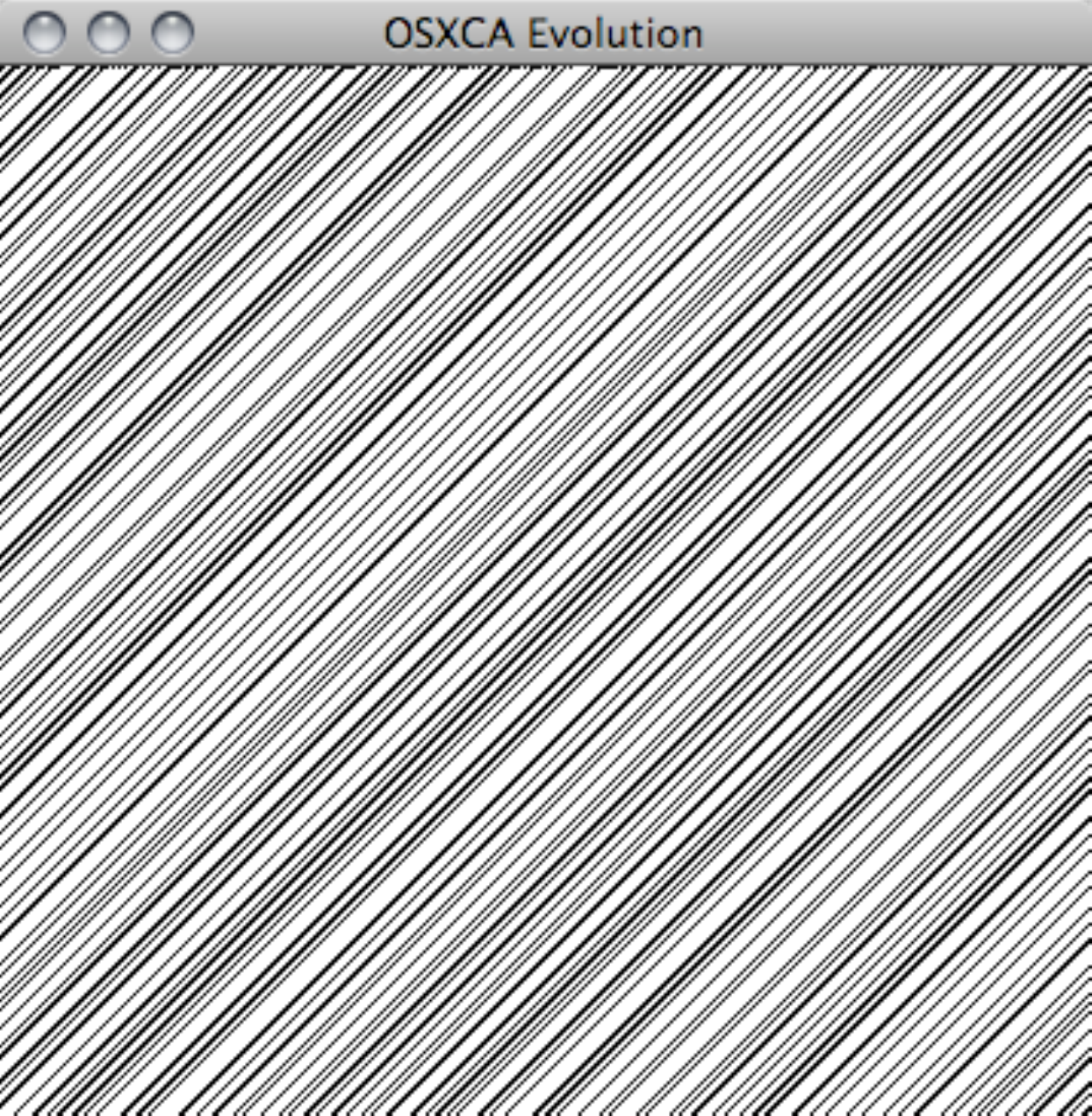}}} 
\subfigure[]{\scalebox{0.47}{\includegraphics{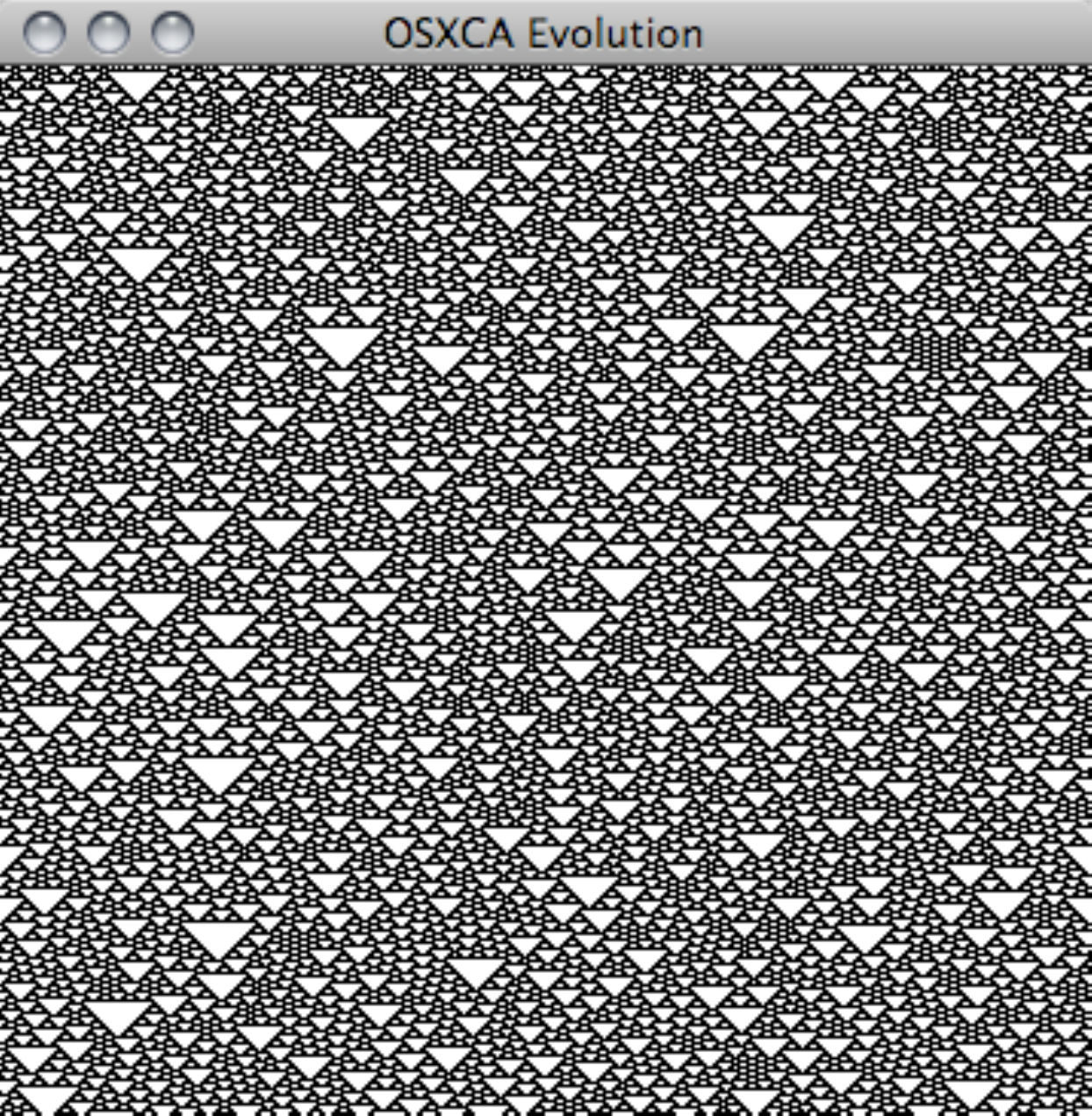}}} 
\subfigure[]{\scalebox{0.47}{\includegraphics{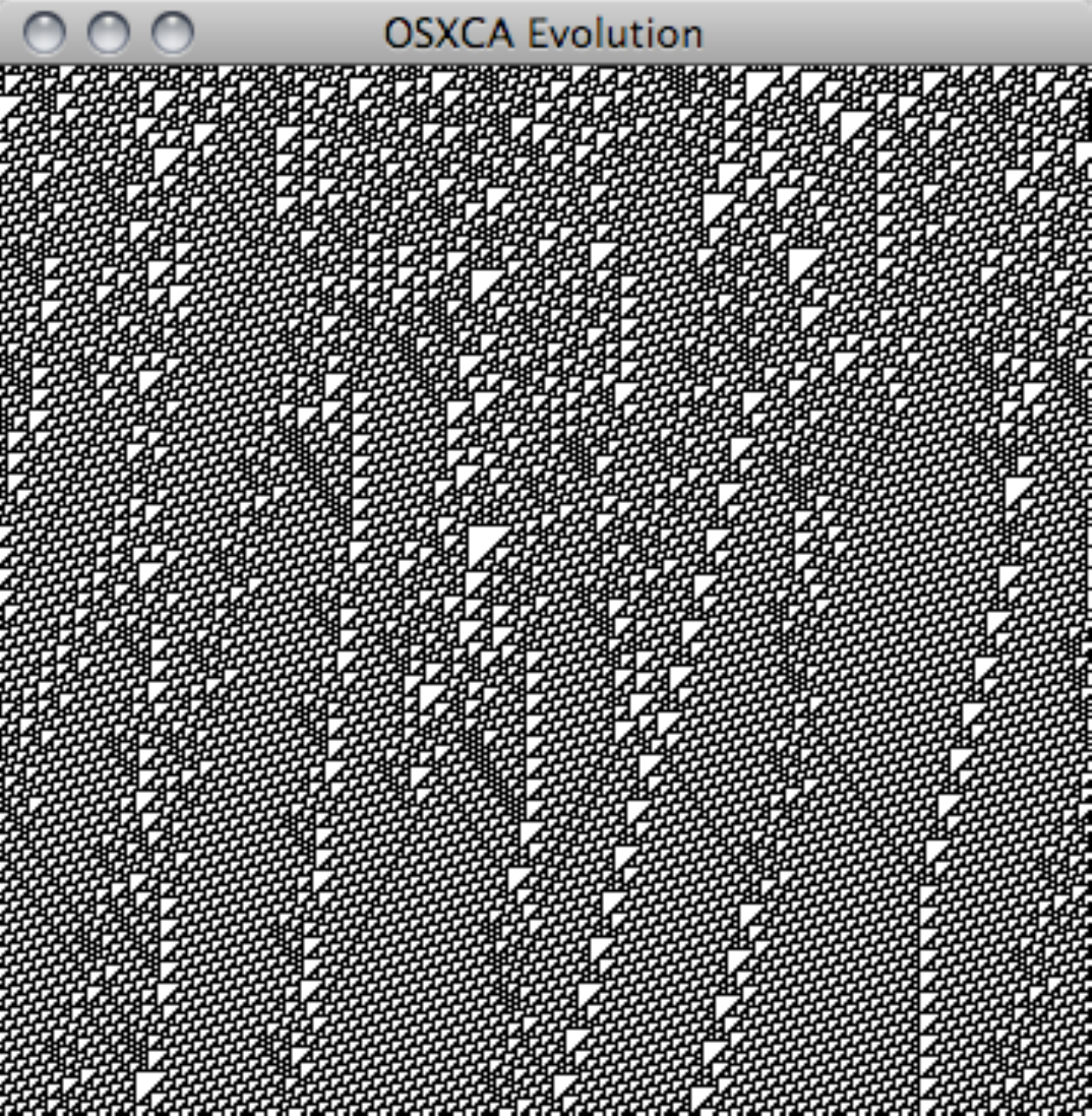}}} \end{center}
\caption{Wolfram's classes represented by ECA rules: (a) Class I - ECA Rule 32, (b) Class II - ECA Rule 10, (c) Class III - ECA Rule 126, (d) Class IV - ECA Rule 110. We have the same initial condition in all these cases, with a density of 50\% for state 0 (white dots) and state 1 (black dots). The evolution space begins with a ring of 358 cells for 344 generations.}
\label{WolframClasses}
\end{figure}

Fig.~\ref{WolframClasses} illustrates Wolfram's classes, focusing on a specific ECA evolution rule (following  Wolfram's notation for ECA \cite{kn:Wolf83}). All evolutions begin with the same random initial condition. Thus, Fig.~\ref{WolframClasses}a displays ECA Rule 32 converging quickly to a homogeneous state, Class I. Figure~\ref{WolframClasses}b displays blocks of cells in state one which evolve periodically showing a leftward shift, Class II. Figure~\ref{WolframClasses}c displays a typical chaotic evolution, where no pattern can be recognised or any limit point identified, Class III. Finally, Fig.~\ref{WolframClasses}d displays the so called complex class or Class IV. Here we see non-trivial patterns emerging in the evolution space. Such patterns possess a defined form and travel along the evolution space. They interact (collide), giving rise to interesting reactions such as annihilations, fusions, solitons and reflections, or they produce new structures. These patterns are referred to as {\it gliders} in the CA literature ('glider' is a widely accepted concept popularised by John Conway through his well-known additive 2D CA, the {\it Game of Life} (GoL) \cite{kn:Gard70}). In Class IV CA we see regions with periodic evolutions and chaos, and most frequently in complex rules the background is 
dominated by stable states, such as in GoL. In such cases---and this is particularly true of the complex ECA Rule 110--the CA can evolve with a periodic background (called ether) where these gliders emerge and live. Gliders in GoL and other CAs such as the 2D Brian's Brain CA \cite{kn:TM87} caught the attention of Christopher Langton, spurring the development of the field of {\it Artificial Life} (AL) \cite{kn:Lang84, kn:Lang86}. 

Since the publication of the paper ``Universality and complexity in cellular automata'' in 1984 \cite{kn:Wolf84}, such classifications have been a much studied and much disputed subject. Wolfram cited several ECA rules as representatives of each class. Despite commenting that (page 31): {\it $k=2, r=1$ cellular automata are too simple to support universal computation.}, in his book ``Cellular Automata and Complexity'' \cite{kn:Wolf94} ECA Rule 110 was granted its own appendix (Table 15, Structures in Rule 110, pages 575--577). It contains specimens of evolutions, including a list of thirteen gliders compiled by Doug Lind, and also presents the conjecture that the rule could be universal. Wolfram writes: {\it One may speculate that the behaviour of Rule 110 is sophisticated enough to support universal computation}. 

An interesting paper written by Karel Culik II and Sheng Yu titled ``Undecidability of CA Classification Schemes'' \cite{kn:CY88, kn:Sut89} discussed the properties of such classes, concluding that: {\it it is undecidable to which class a given cellular automaton belongs} (page 177). Indeed, in 1984 Wolfram \cite{kn:Wolf84} commented (page 1): {\it The fourth class is probably capable of universal computation, so that properties of its infinite time behaviour are undecidable}. Actually, we can see that no effective algorithm exists that is capable of deciding whether a CA is complex or universal, and so far only a few discovered (as opposed to constructed) cellular automata have been proven to be capable of universal computation (notably Wolfram's Rule 110 and Conway's Game of Life). However some techniques offer suitable approximations for finding certain sets of complex, though not necessarily universal rules (though under Wolfram's PCE they would be, c.f. Section \ref{finalremarks}). 

In ``Local structure theory for cellular automata'' \cite{kn:GVK87} Howard Gutowitz has developed a statistical analysis. An interesting schematic diagram conceptualising the umbral of classes of CA was offered by Wentian Li and Norman Packard in ``The Structure of the Elementary Cellular Automata Rule Space'' \cite{kn:LP90}. Pattern recognition and classification has been examined in ``Toward the classification of the patterns generated by one-dimensional cellular automata'' \cite{kn:AN86}. An extended analysis by Andrew Adamatzky under the heading ``Identification of Cellular Automata'' in \cite{kn:Ada94} considered the problem of how, given a sequence of configurations of an unknown cellular automaton, one may reconstruct its evolution rules. A recent special issue dedicated to this problem focuses on some theoretical and practical results.\footnote{Special issue ``Identification of Cellular Automata'', {\em Journal of Cellular Automata} {\bf 2(1)}, 1--102, 2007. \url{http://www.oldcitypublishing.com/JCA/JCAcontents/JCAv2n1contents.html}} Klaus Sutner has discussed this classification and also the principle of computational equivalence in ``Classification of Cellular Automata'' \cite{kn:Sut09}, with an emphasis on Class IV or computable CA. An interesting approach involving an additive 2D CA was described in David Eppstein's classification scheme \cite{kn:Epp99}\footnote{For a discussion see Tim Tyler's CA FAQ at \url{http://cafaq.com/classify/index.php}, and more recently, a compression-based technique inspired by algorithmic information theory has been advanced\cite{kn:Zen10} that offers a powerful method for identifying complex CA and other complex systems }. 

We will discuss two practical and two theoretical studies that distinguish such classes and explore some properties of computing CA rules. Among the topics we want to explore is the feasibility of using extended analog computers (EAC) \cite{kn:Mills08} for CA construction, in order to obtain unconventional computing models \cite{kn:Ada02, kn:Ada01}. In this classification, Class IV is of particular interest because the rules of the class present non-trivial behaviour, with a rich diversity of patterns emerging, and non-trivial interactions between gliders, plus mobile localizations, particles, or fragments of waves. This feature was useful in implementing a register machine in GoL \cite{kn:BCG82} to determine its universality. First we survey some of the approximations that allow the identification of complex properties of CA and other systems.

\subsection{Mean field approximation}

Mean field theory is a well-known technique for discovering the statistical properties of CA without analysing the evolution space of individual rules. It has been used extensively by Gutowitz in \cite{kn:Guto89}. The method assumes that states in $\Sigma$ are independent and do not correlate with each other in the local function $\varphi$. Thus we can study probabilities of states in a neighbourhood in terms of the probability of a single state (the state in which the neighbourhood evolves), and the probability of a neighbourhood would be the product of the probabilities of each cell in it.

Harold V. McIntosh in \cite{kn:Mc90} presents an explanation of Wolfram's classes using a mixture of probability theory and de Bruijn diagrams\footnote{The de Bruijn diagrams have been culled from Masakazu Nasu's 1978 work on tessellation automata \cite{kn:Nasu78}. Wolfram himself has explored some of this in  \cite{kn:Wolf84a}, later thoroughly analysed by McIntosh \cite{kn:Mc91, kn:Mc09}, Sutner \cite{kn:Sut91}, Burton Voorhes \cite{kn:Voor96, kn:Voor06}, and, particularly, exploited to calculate reversible 1D CA using de Bruijn diagrams derived from the Welch diagrams by Seck-Tuoh-Mora in \cite{kn:SCM05, kn:SMM06}}, resulting in a classification based on the mean field theory curve:

\begin{myitemize}
\item Class I: monotonic, entirely on one side of diagonal;
\item Class II: horizontal tangency, never reaches diagonal;
\item Class IV: horizontal plus diagonal tangency, no crossing;
\item Class III: no tangencies, curve crosses diagonal.
\end{myitemize}

For the one-dimensional case, all neighbourhoods are considered, as follows:

\begin{equation}
p_{t+1}=\sum_{j=0}^{k^{2r+1}-1}\varphi_{j}(X)p_{t}^{v}(1-p_{t})^{n-v}
\label{MFp1D}
\end{equation}

\noindent such that $j$ indexes every neighbourhood, $X$ are cells $x_{i-r}, \ldots, x_{i}, \ldots, x_{i+r}$, $n$ is the number of cells in every neighbourhood, $v$ indicates how often state `1' occurs in $X$, $n-v$ shows how often state `0' occurs in the neighbourhood $X$, $p_{t}$ is the probability of a cell being in state `1', and $q_{t}$ is the probability of a cell being in state `0'; i.e., $q=1-p$. For mean field theory in other lattices and dimensions, please consult \cite{kn:Guto87a, kn:Guto89a}.

\subsection{Basins of attraction approximation}
Andrew Wuensche, together with Mike Lesser, published a landmark book entitled ``The Global Dynamics of Cellular Automata'' in 1992 \cite{kn:WL92} which contained a very extended analysis of attractors in ECA. Wolfram himself had explored part of these cycles in ``Random Sequence Generation by Cellular Automata'' \cite{kn:Wolf86}, as had McIntosh in ``One Dimensional Cellular Automata'' \cite{kn:Mc09}. Notably, Stuart Kauffman in his book ``The Origins of Order: Self-Organization and Selection in Evolution'' \cite{kn:Kau93} applies basins of attraction to sample random Boolean networks (RBN) in order to illustrate his idea that RBN constitute a model of the gene regulatory network, and that cell types are attractors. The best description of such an analysis is to be found in \cite{kn:Wue98}. 

A basin (of attraction) field of a finite CA is the set of basins of attraction into which all possible states and trajectories will be organized by the local function $\varphi$. The topology of a single basin of attraction may be represented by a diagram, the {\it state transition graph}. Thus the set of graphs composing the field specifies the global behaviour of the system \cite{kn:WL92}. 

Generally a basin can also recognize CA with chaotic or complex behaviour using prior results on attractors \cite{kn:WL92}. Thus, Wuensche says that Wolfram's classes can be represented as a {\it basin classification} \cite{kn:WL92}, as follows:

\begin{myitemize}
\item Class I: very short transients, mainly point attractors (but possibly also periodic attractors), very high in-degree, very high leaf density (very ordered dynamics);
\item Class II: very short transients, mainly short periodic attractors (but also point attractors), high in-degree, very high leaf density;
\item Class IV: moderate transients, moderate-length periodic attractors, moderate in-degree, very moderate leaf density (possibly complex dynamics);
\item Class III: very long transients, very long periodic attractors, low in-degree, low leaf density (chaotic dynamics).
\end{myitemize}

\subsection{Compressibility approximation}
\label{compressibility} 

More recently, a compression-based classification of CA (and other systems) was proposed in \cite{kn:Zen10}, based on concepts from algorithmic complexity. The technique is based on the notion of asymptotic behaviour, and unlike the mean field theory technique, it analyses the statistical properties of CA by looking at the evolution space of individual rules. The method produced the following variation of Wolfram's classification \cite{kn:ZenAISB}. 

\begin{myitemize}
\item Class I: highly compressible evolutions for any number of steps;
\item Class II: highly compressible evolutions for any number of steps;
\item Class III: the lengths of compressed evolutions asymptotically converge to the uncompressed evolution lengths;
\item Class IV: the lengths of compressed evolutions asymptotically converge to the uncompressed evolution lengths.
\end{myitemize}

The first problem we face is that the four classes seem to give way to two (Classes I and II and Classes III and IV are grouped). We will briefly see how algorithmic information theory helps to separate them again, using the concept of asymptotic behaviour advanced in \cite{kn:Zen10,kn:Zen12}.

\begin{figure}
\begin{center}
\subfigure[]{\scalebox{0.29}{\includegraphics{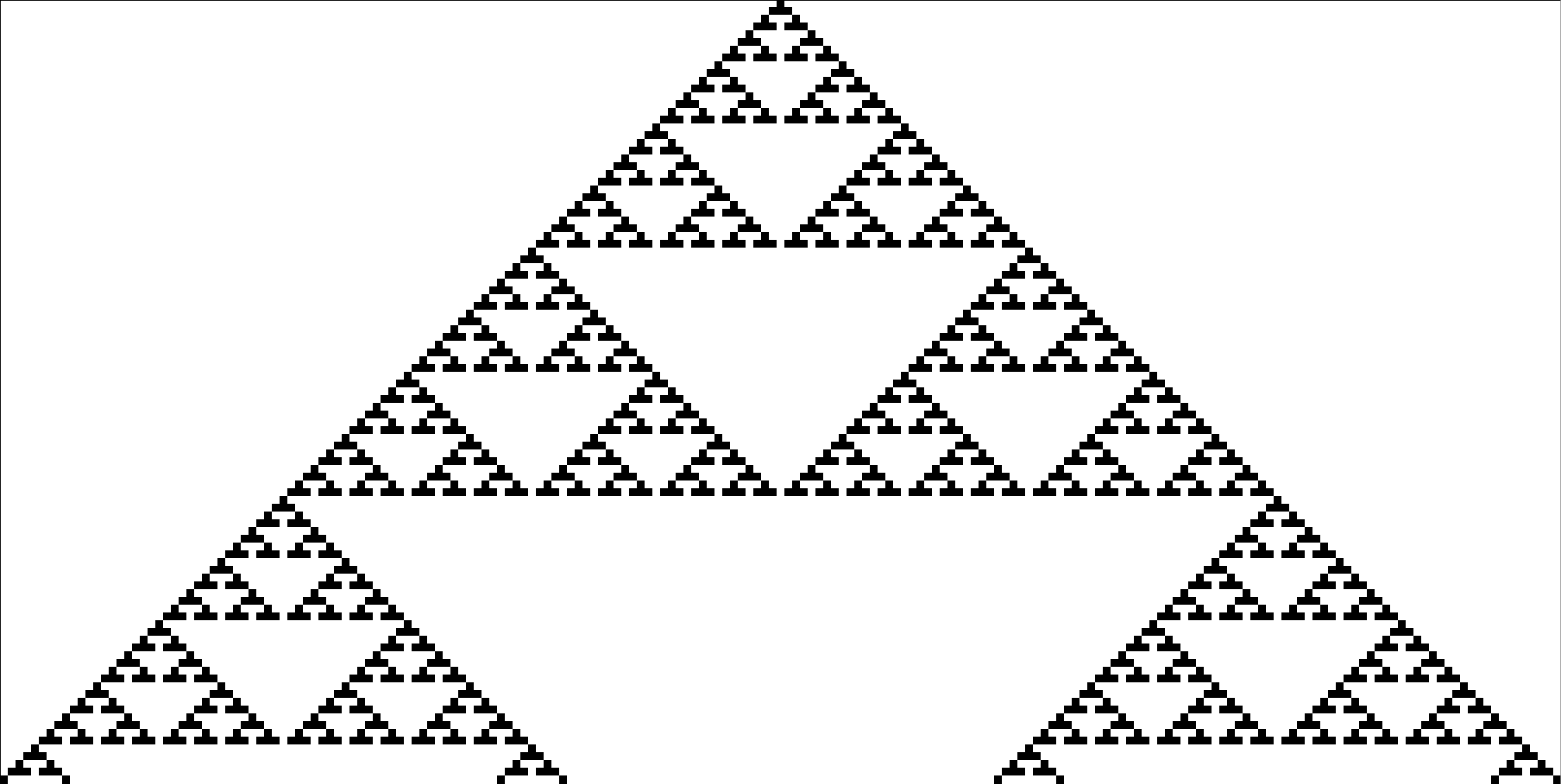}}} 
\subfigure[]{\scalebox{0.29}{\includegraphics{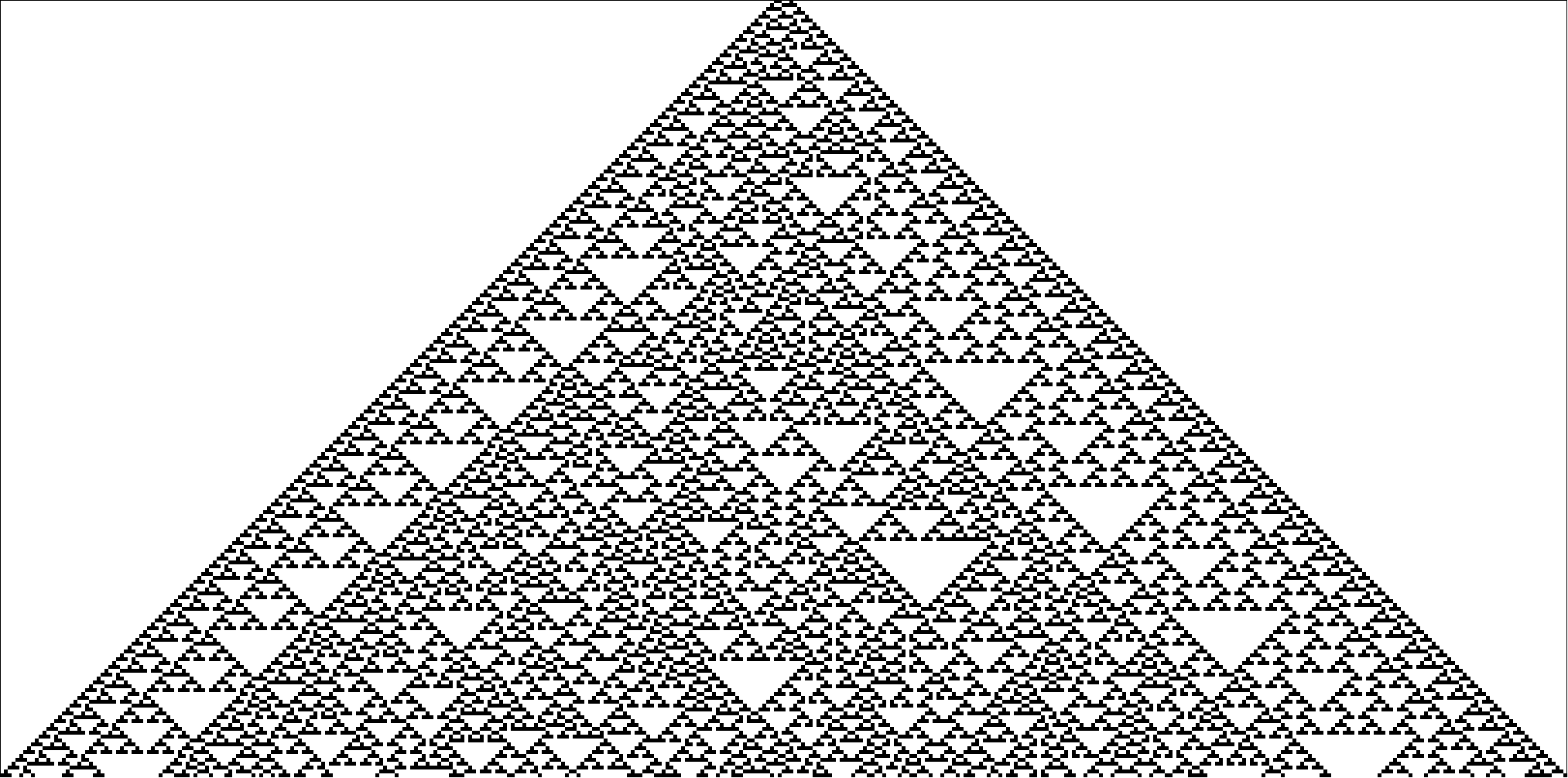}}}
\subfigure[]{\scalebox{0.29}{\includegraphics{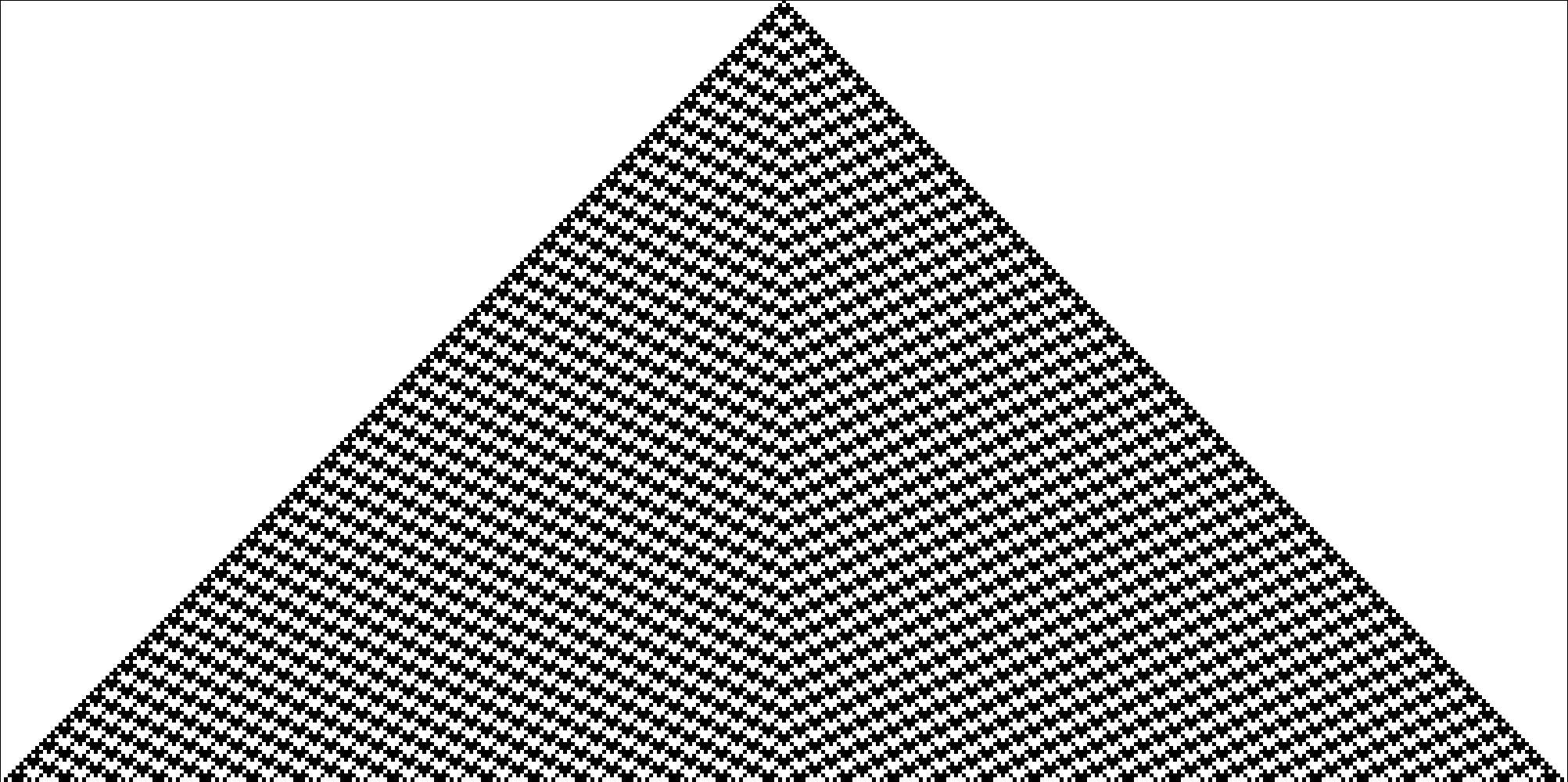}}} 
\subfigure[]{\scalebox{0.29}{\includegraphics{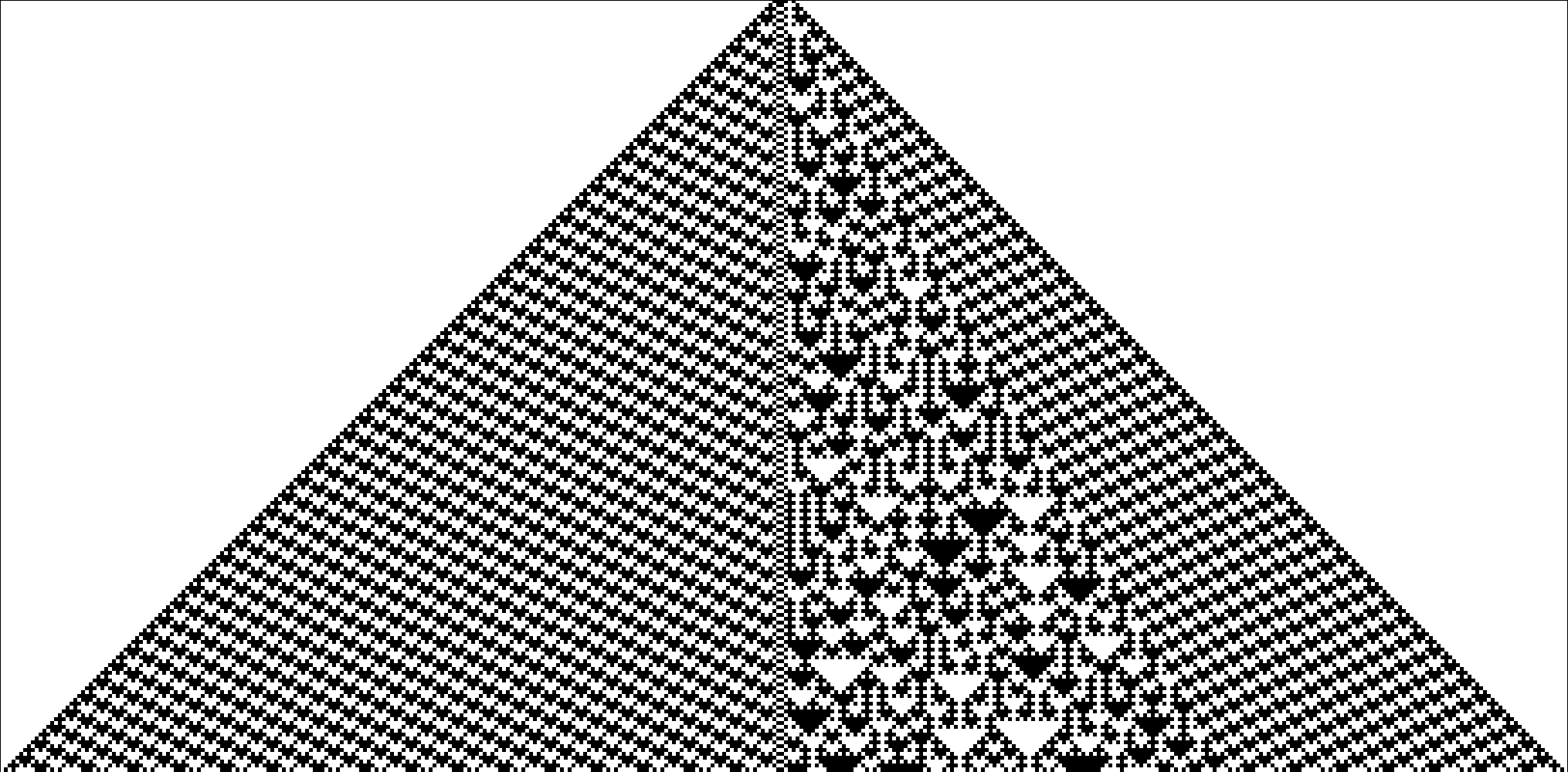}}}\end{center}
\caption{To which of Wolfram's Classes do these two ECAs (Rule 22 and Rule 109) belong? (a) Wolfram's ECA Rule 22 starting from a single black cell, (b) Rule 22 starting from another initial configuration (11001), (c) Wolfram's ECA Rule 109 starting from a single black cell, (d) The same Rule 109 starting from another initial configuration (111101).}
\label{zenilfig1}
\end{figure}

The investigation in \cite{kn:Zen10} provides one of several ways to address the criticisms directed at Wolfram's original classification. In the experiments that led Wolfram to propose his classification he started the systems with a ``random" initial configuration as a way to \emph{sample} the behaviour of a system and circumvent the problem of having to choose a particular initial configuration to map a system to its possible class of behaviour. The problem resides in the fact that a CA, like any other dynamical system, may have phase transitions, behaving differently for different initial configurations (the question is ultimately undecidable (see \cite{kn:CY88})). The chances of having a CA display its \emph{average} behaviour (that is, its behaviour for \emph{most} initial configurations) are greater when taking a ``random" initial configuration, if one assumes that there is no bias towards any particular region of the possible enumerations of initial configurations (consider the behaviour of a CA starting from one initial configuration versus another (see Figures~\ref{zenilfig1}). One can even enumerate initial configurations according to how a system behaves, hence artificially biasing the result for a potentially arbitrary number of initial conditions, making a system assume a particular appearance for an arbitrary length of time). 

If, for example, one enumerates binary initial configurations in Wolfram's tradition (in a decimal-to-binary code), one has to assume that any segment of the enumeration yields the same average behaviour when using a ``random" initial configuration. In a word, if a behaviour occurs more often when a ``random" initial configuration is used, the chances are greater that it will represent the ``average" behaviour of the CA. Yet this does not solve the problem of a system that may behave in a completely unprecedented fashion for a runtime or a set of initial configurations not explored before. In \cite{kn:Zen10}, however, the idea is to introduce the concept of \emph{a posteriori} asymptotic behaviour of a CA inspired by techniques in dynamical systems but to the actual evolutions of a system (in order to make the measure applicable to natural systems), that is both how a CA behaves over time and for an initial segment of initial configurations. 

But in order to capture the``natural" behaviour of a CA one has first to devise a better way to enumerate initial configurations than the traditional decimal-to-binary method, so as to overcome the problem of introducing artificial phase transitions from the input. Consider the decimal-to-binary enumeration where the initial input with traditional number 31 in decimal, converted to binary, is the initial input 11111 for the CA, as contrasted with the initial input 32 in decimal, 100000 in binary. One shouldn't then be surprised to see a qualitative change in the behaviour of a system arising from such a non-natural change from one initial input to the \emph{next one}. However, if one guarantees that only a bit will change from one initial configuration to the next, one can avoid such cases. This is what a Gray (or Gray-Gros) code based enumeration of initial configurations allows, as explained in \cite{kn:Zen10}.

 This treatment permits the definition of a compression-based phase transition coefficient capturing the \emph{asymptotic behaviour} of a system \cite{kn:Zen10,kn:Zen12}, which in turn allows us to separate the collapsed classes and even advance a different and alternative classification, based on the sensitivity of a CA to its initial conditions \cite{kn:Zen10}, which has also been conjectured to be related to the system's ability to transfer information, and ultimately to its computing abilities, particularly as these relate to Turing universal computation (see \cite{kn:Zen12}). This approach does not solve the problem of a system that behaves in a  qualitatively different manner after a certain number of initial input configurations or after a certain period of time (the same problem encountered when devising the original classification), which is not a problem of method, but is instead related to the general problem of induction and of reachability (hence to undecidability in general). Nonetheless it does address the problem of a reasonable definition of the ``average behaviour" of a system (in this case a CA) under the same assumptions made for other enumerations (viz. that enumerations, especially natural ones, have no distinct regions where a system starts behaving in a completely different fashion, making it impossible to talk about the convergence in behaviour of a system). Wolfram's classes can once again be separated using the compression-based approach in combination with the following classification \cite{kn:ZenAISB}, derived from a phase transition coefficient presented in \cite{kn:Zen10}:

\begin{myitemize}
\item Class I: insensitivity to initial configurations, inability to transfer information other than isolated bits;
\item Class II: sensitivity to initial conditions, ability to transfer some information;
\item Class III: insensitivity to initial configurations, inablility to transfer information, perhaps due to lack of (evident means of) control;
\item Class IV: sensitivity to initial conditions, ability to transfer some information.
\end{myitemize}

One can only understand how Classes I and III can now be together in this classification on the basis of the qualitative treatment explained above. In other words, when one changes the initial configuration of a system in either of these two classes (I and III) the system's behaviour remains the same (each evolution is equally compressible), and it is therefore considered unable to or inefficient at transferring information or programming a CA to perform (universal) computation. On the other hand, classes II and IV are better at transferring information, even if they may do so in different ways. This classification tells us that classes II and IV are more sensitive to initial configurations (e.g. Wolfram's ECA rule 22, considered to belong to Class II, and Wolfram's ECA Rule 110 belonging to Class IV). Another way to show how members of Classes I and III may belong to the same class in the second classification is through another interesting and useful concept, the concept of \emph{heat} (c.f. Section \ref{heat}).

 Together, the compression-based classifications capturing different behaviours of the systems capture other intuitive notions that one would expect from Wolfram's original classification. The values for ECA calculated in \cite{kn:Zen10}  yielded results that also suggest that one may be able to relate these measures to universality through the definition of Class IV, as given above (see \cite{kn:Zen12}).

\section{Universal CA Class IV versus Class III}

Culik II and Yu have demonstrated \cite{kn:CY88} that whether a CA belongs to Class IV is undecidable.  Nevertheless, some approximations have been developed, with interesting results. The use of genetic programming by Melanie Mitchell, Rajarshi Das, Peter Hraber, and James Crutchfield \cite{kn:MHC93, kn:DMC94} to obtain sets of rules with particles and computations is a case in point. As indeed is Emmanuel Sappin's calculation of a non-additive universal 2D CA with a genetic algorithm, the {\it R rule} \cite{kn:SBC04, kn:SBC07}. However, the use of evolutionary techniques has been limited to a small portion of complex CA with few states and small configurations. Up to now, brute force programming has been necessary to obtain monsters of complex patterns in huge spaces, as Eppstein shows in \cite{kn:Epp02}.

\begin{figure}
\begin{center}
\subfigure[]{\scalebox{0.27}{\includegraphics{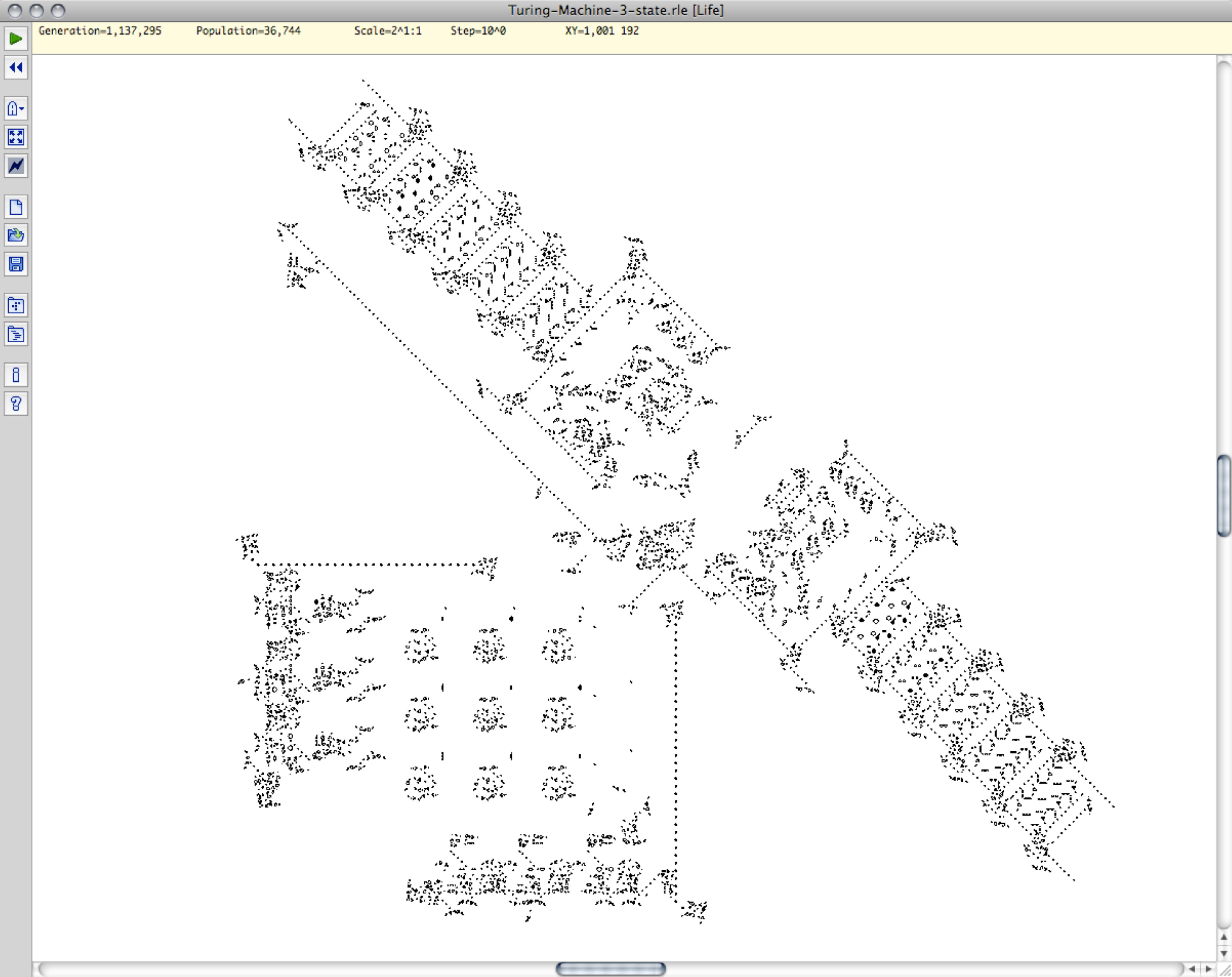}}} \hspace{0.8cm}
\subfigure[]{\scalebox{0.35}{\includegraphics{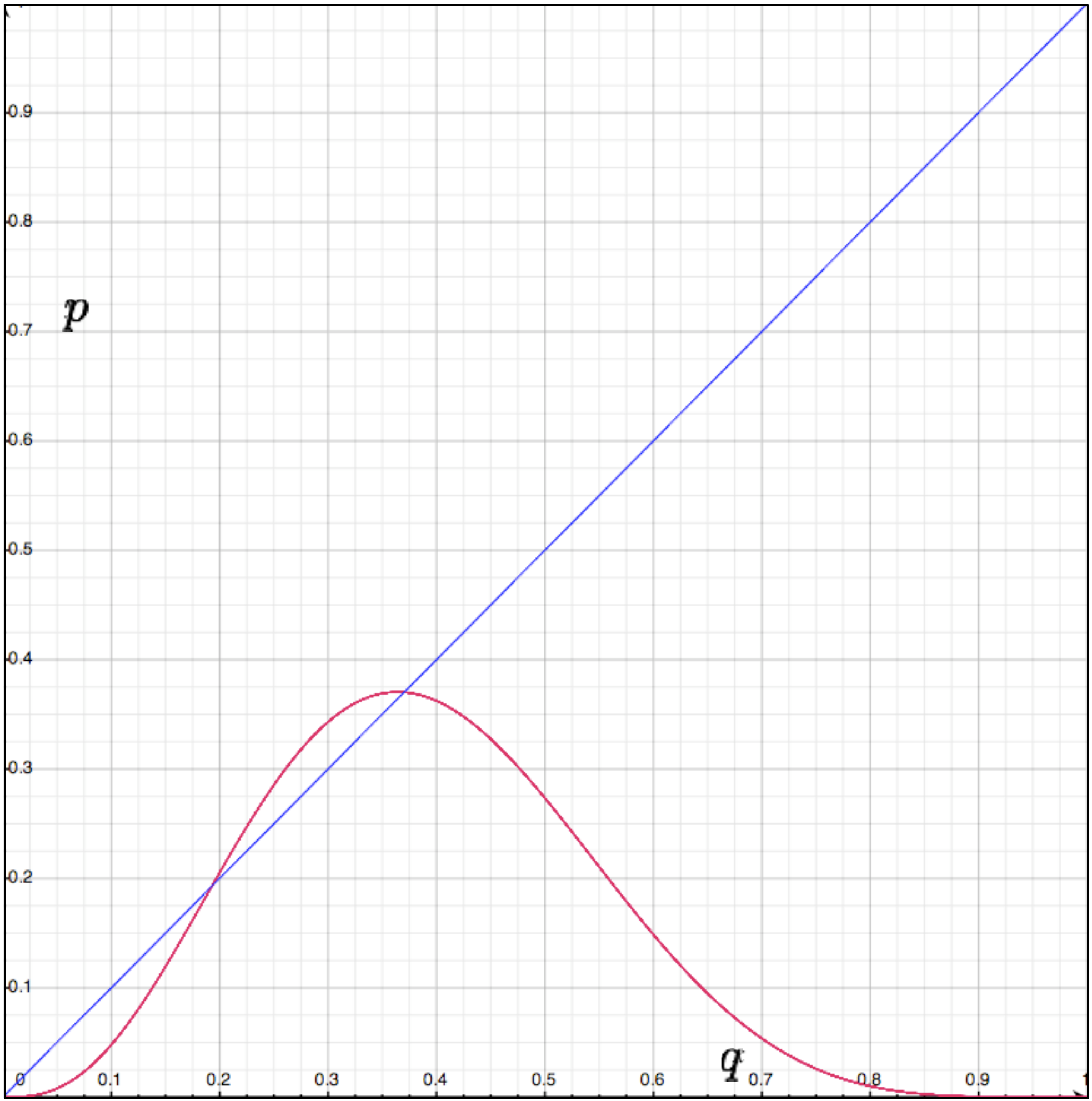}}}
\end{center}
\caption{(a) A 3-state, 3-symbol Turing machine in GoL by Rendell \cite{kn:Ren02, kn:Ren11}, (b) its mean field curve.}
\label{GoL_TM_MF}
\end{figure}

\subsection{The Game of Life: Class IV}

The most popular 2D CA is certainly Conway's Game of Life (GoL), a binary 2D additive CA, first published in Martin Garden's column in {\it Scientific American} \cite{kn:Gard70}. GoL can be represented as $R(2,3,3,3)$, or typically, as the  $B3/S23$ rule.\footnote{An excellent forum on GoL is ``LifeWiki'' \url{http://conwaylife.com/wiki/index.php?title=Main_Page}. To complement this, you may consult  ``The Game of Life Sites'' \url{http://uncomp.uwe.ac.uk/genaro/Cellular_Automata_Repository/Life.html}.} In 1982, Conway proved that GoL was universal by developing a register machine working with gliders, glider guns, still life and oscillator collisions \cite{kn:BCG82}. However, such universality was completed by Paul Rendell's demonstration in 2000 that involved implementing a 3-state, 3-symbol Turing machine in GoL \cite{kn:Ren02, kn:Ren11}. The machine duplicates a pattern of 1's within two 1's on the tape to the right of the reading position, running 16 cycles to stop with four 1's on the tape. A snapshot of this implementation is provided in Fig.~\ref{GoL_TM_MF}a. For details about each part and about the functionality of this machine please visit ``Details of a Turing Machine in Conway's Game of Life'' \url{http://rendell-attic.org/gol/tmdetails.htm}. 

GoL is a typical Class IV CA evolving with complex global and local behaviour. In its evolution space we can see a number of complex patterns which emerge from different configurations. Gol has been studied since 1969 by Conway, and William Gosper of MIT's Artificial Life research group has taken a strong interest in it. The tradition of GoL research is very much alive, with today's GoL researchers discovering new and very complex constructions by running complicated algorithms. Just last year, GoL celebrated its 40th anniversary. The occasion was marked by the publication of the volume ``Game of Life Cellular Automata'' \cite{kn:Ada10}, summarising a number of contemporary and historical results in GoL research as well as work on other interesting Life-like rules. 

According to mean field theory, $p$ is the probability of a cell's being in state `1' while $q$ is its probability of its being in state `0' i.e., $q=1-p$, and the {\it mean field equation} represents the neighbourhood that meets the requirement for a live cell in the next generation \cite{kn:Mc90}. As we have already seen, horizontal plus diagonal tangency, not crossing the identity axis (diagonal), and the marginal stability of the fixed point(s) due to their multiplicity indicates Wolfram's Class IV \cite{kn:Guto89}, or complex behaviour. Hence, we will review the global behaviour of GoL using mean field theory. Figure~\ref{GoL_TM_MF}b shows the mean field curve for GoL, with polynomial:

$$
p_{t+1}=28p^3_tq^5_t(2p_t+3q_t).
$$

The origin is a stable fixed point, while the unstable fixed point $p=0.2$ represents the fact that densities around 20\% induce complex behaviour for configurations in such a distribution. $p=0.37$ is the maximum stable fixed point where GoL commonly reaches global stability inside the evolution space.\\

\begin{figure}
\begin{center}
\scalebox{1}{\includegraphics{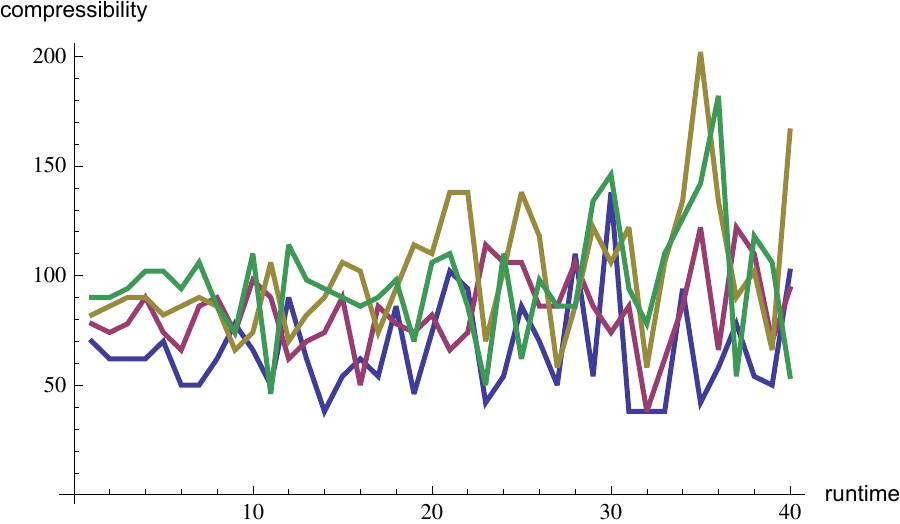}}%
\caption{\label{figcomp} Lossless compressing (with the DEFLATE algorithm) the evolution of the Game of Life starting from 4 different random initial conditions from 1 to 40 steps (generations) shows that the behaviour of the system remains relatively complex, given that none led to substantial compressibility, as would have been the case if patterns die out after a period of time.}
\end{center}
\end{figure}

In \cite{kn:ZenJETAI} a compression-based coefficient for GoL was calculated, showing that, as expected, it exhibits a high degree of \emph{variability} (see Fig. \ref{figcomp}) and \emph{programmability}. This is in agreement with the known fact that GoL is capable of universal computation, and hence supports the notion discussed in \cite{kn:Zen12} that sensitivity to initial configurations and rate of information transmission is deeply connected to (Turing) universality and can be measured by metrics inspired in approximations to algorithmic complexity.

\subsection{Life-like rule $B35/S236$: Class III}

\begin{figure}
\begin{center}
\subfigure[]{\scalebox{0.35}{\includegraphics{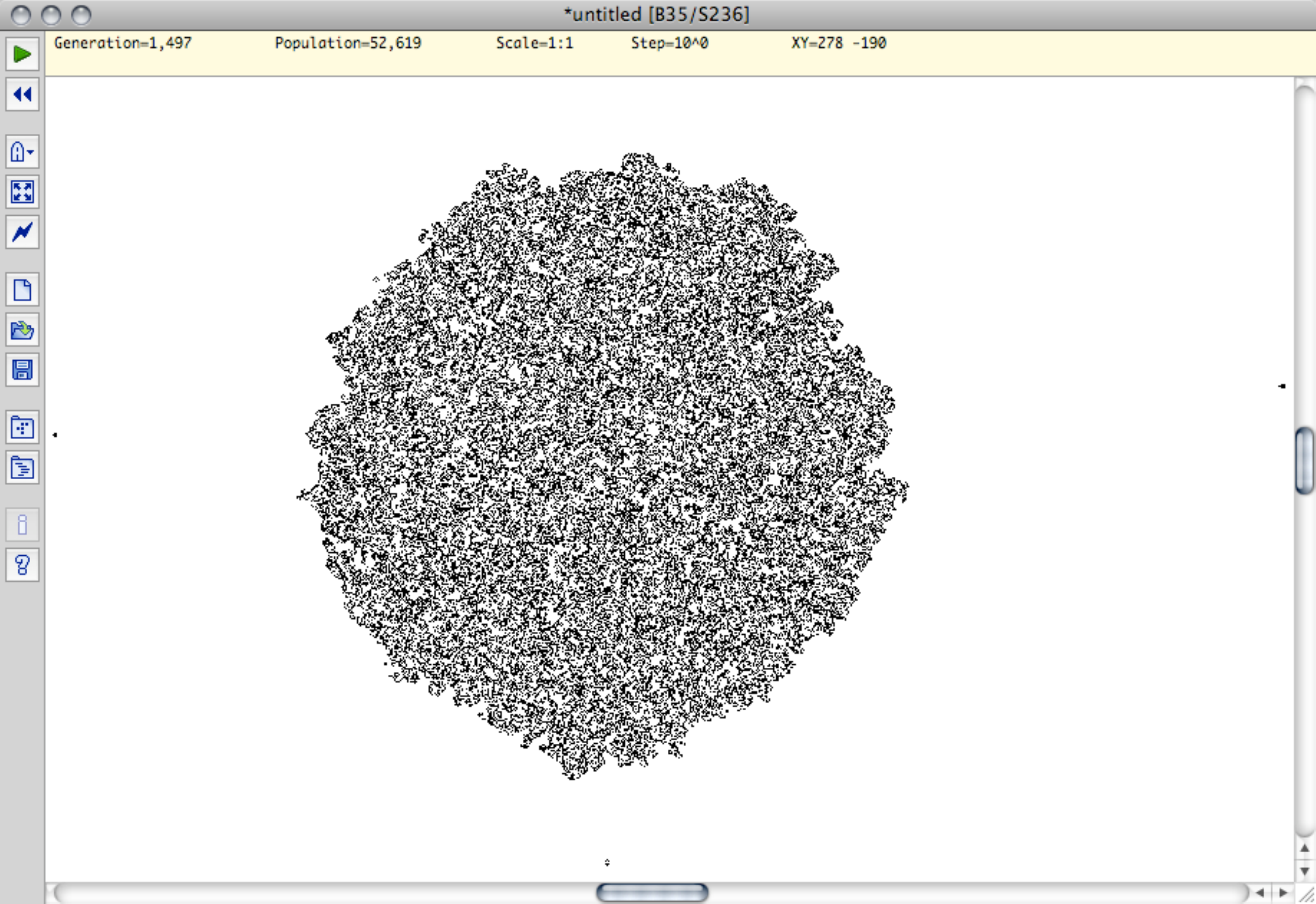}}} \hspace{0.8cm}
\subfigure[]{\scalebox{0.35}{\includegraphics{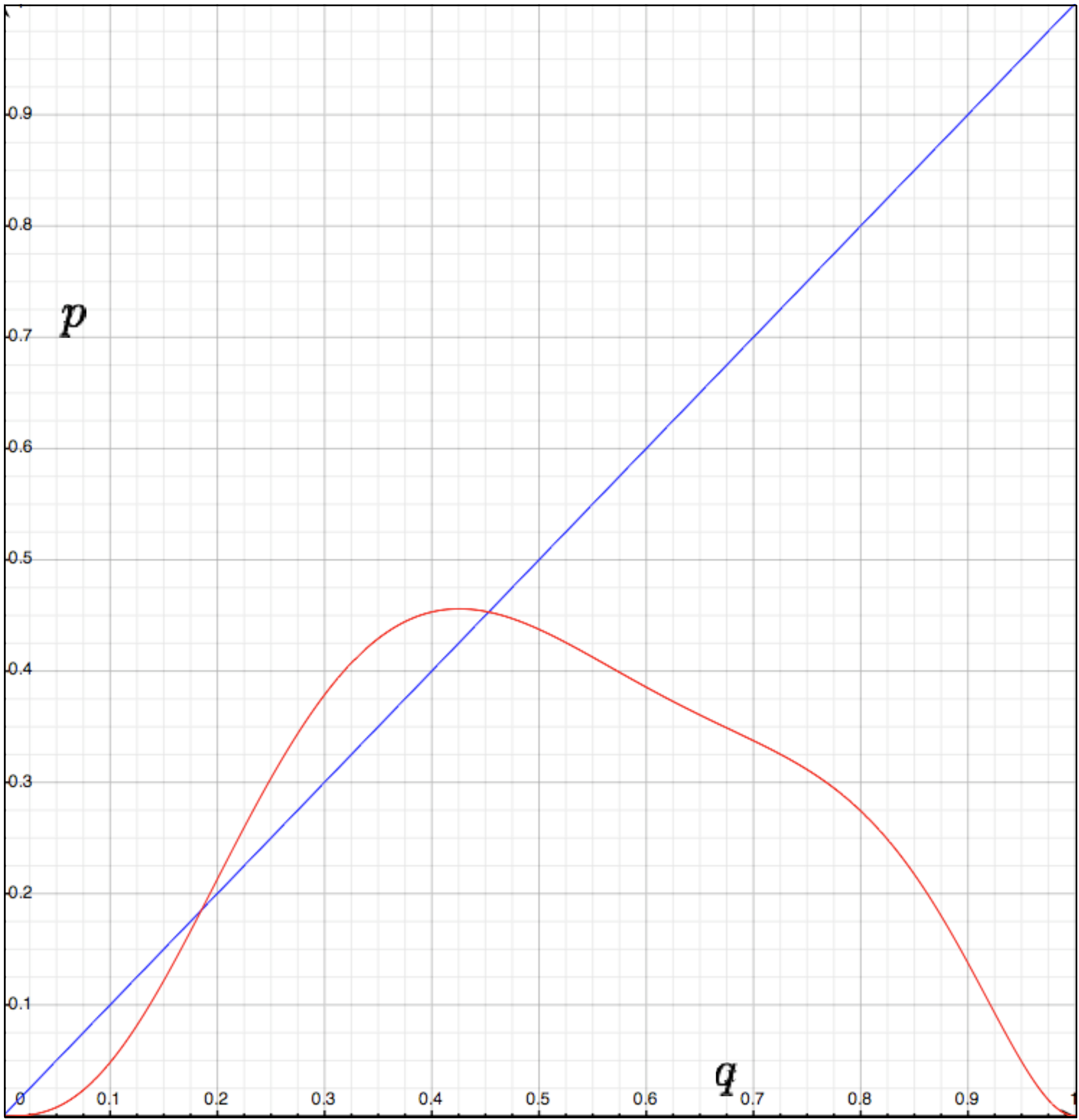}}}
\end{center}
\caption{(a) Evolution starting from an L-pentomino in Life-like CA $B35/S236$, (b) its mean field curve.}
\label{B35_S236}
\end{figure}

The Life-like CA evolution rule $B35/S236$ was proposed by Eppstein and Dean Hickerson as a chaotic CA with sufficient elements for developing universality. Details about these computable elements are available at \url{http://www.ics.uci.edu/~eppstein/ca/b35s236/construct.html}. The family of gliders and other complex constructions in this rule can be found at \url{http://www.ics.uci.edu/~eppstein/ca/b35s236/}.

 The $B35/S236$ automaton commonly evolves chaotically. Figure~\ref{B35_S236}a displays a typical chaotic evolution starting from an L-pentomino configuration; after 1,497 generations  there is a population of 52,619 live cells. Here we see how a few gliders emerge from chaos and then quickly escape, although the predominant evolution over a long period is chaotic.

 Figure~\ref{B35_S236}b shows the mean field curve for CA $B35/S236$, with polynomial:

$$
p_{t+1}=28p^3_tp^2_t(p^4_t+2p_tq^3_t+2p^2_tq^2_t+3q^4_t).
$$

The origin is a stable fixed point (as in GoL) which guarantees the stable configuration in zero, while the unstable fixed point $p=0.1943$  (again very similar to GoL) represents densities where we could find complex patterns emerging in $B35/S236$. $p=0.4537$ is the maximum stable fixed point at which $B35/S236$ commonly reaches global stability.

 This way, $B35/S236$ preserves the diagonal tangency between a stable and an unstable fixed point on its mean field curve. But although its values are close to those of GoL, CA $B35/S236$ has a bigger population of live cells, which is not a sufficient condition for constructing reliable components, especially from unreliable organisms. One of most important von Neumann's feature constructing his universal 29-states was that of universality \cite{kn:von66} but it was not long after that the property of universality was found to be not completely necessary in order to be able to design (or find) a system (a CA) capable of robust self-reproduction, such as Langton's CA known as Langton's Loops.

\subsection{Life-like rule Seeds and the Diffusion Rule: Class III}
There is a special case of a Life-like CA that was originally reported by Brian Silverman and named by Mirek W\'ojtowicz as {\it Seeds} (Life-like rule $B2$).\footnote{Cellular Automata rules lexicon. Family Life: \url{http://psoup.math.wisc.edu/mcell/rullex_life.html}.} Another study presented the so called {\it Diffusion Rule} (Life-like rule $B2/S7$) \cite{kn:MAM10}.\footnote{Diffusion Rule home page: \url{http://uncomp.uwe.ac.uk/genaro/Diffusion_Rule/diffusionLife.html}} The two rules behave identically, as do the rules $B2/S67$, $B2/S6$ and $B2/S5$, which gradually preserve/lose a significant number of complex structures. 

Although the Seeds CA was widely studied, only a number of unreported basic gliders and small glider guns were presented in \cite{kn:MAM10}. Interestingly, $B2/S7$ shows a combination of chaos (dominant evolution) and very stable configurations, including large histories of evolutions. 

Such automata have an amazing number of gliders, puffer trains, avalanches, and glider guns as compared to many other Life-like rules. However, stable configurations cannot be designed. 

\begin{figure}[th]
\centering
\includegraphics[width=1.0\textwidth]{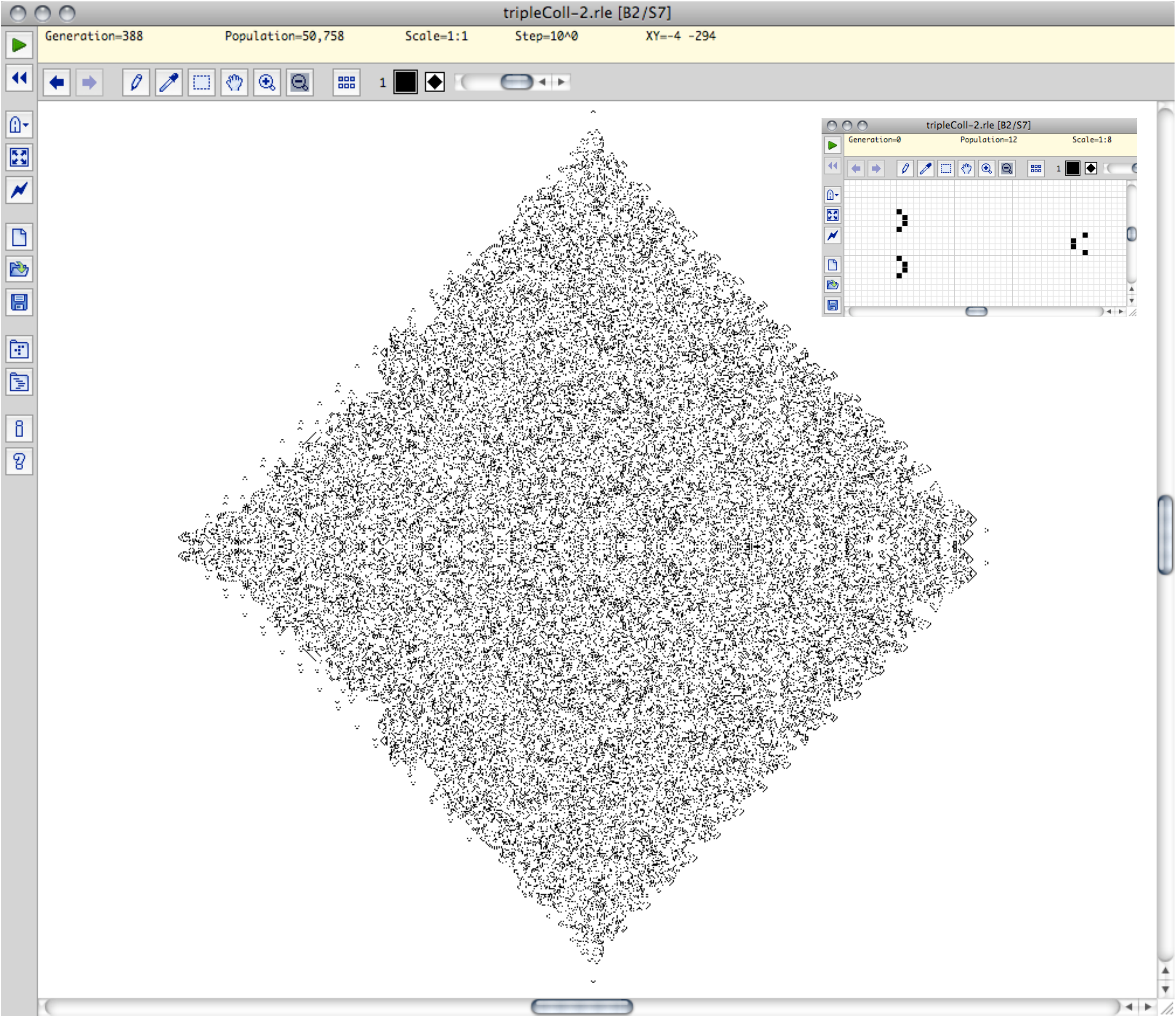}
\caption{Chaos and gliders emerging in the Diffusion Rule from a triple collision between three basic glides.}
\label{tripleCollisions_DR}
\end{figure}

Figure~\ref{tripleCollisions_DR} displays a typical chaotic evolution arising in the Diffusion Rule CA from a triple collision of three basic gliders (initialised at the top right corner of the figure). Thus, the resultant dynamic is quickly dominated by generally chaotic behaviour. Nevertheless, we can see a number of gliders and complex patterns emerging and traveling at the same velocity.

The Diffusion Rule displays complex patterns useful for developing computable devices. Some elements have been developed using these patterns in \cite{kn:MAM10, kn:Ada10}. A problem with this automaton is that the computation must be designed on an infinite space using movable components, just like an {\it Extended Analog Computer} \cite{kn:Mills08}.

 Glider guns guarantee a constant flow of bits and their interactions induce the computation. A very simple memory device was designed in the Diffusion Rule-- between a basic glider and an oscillator-- that can also produce an asynchronous {\sc xnor} and {\sc xor} gate, and it is possible to implement a {\sc fanout} gate as well (for details please see \cite{kn:MAM10}).

\begin{figure}
\begin{center}
\subfigure[]{\scalebox{0.3}{\includegraphics{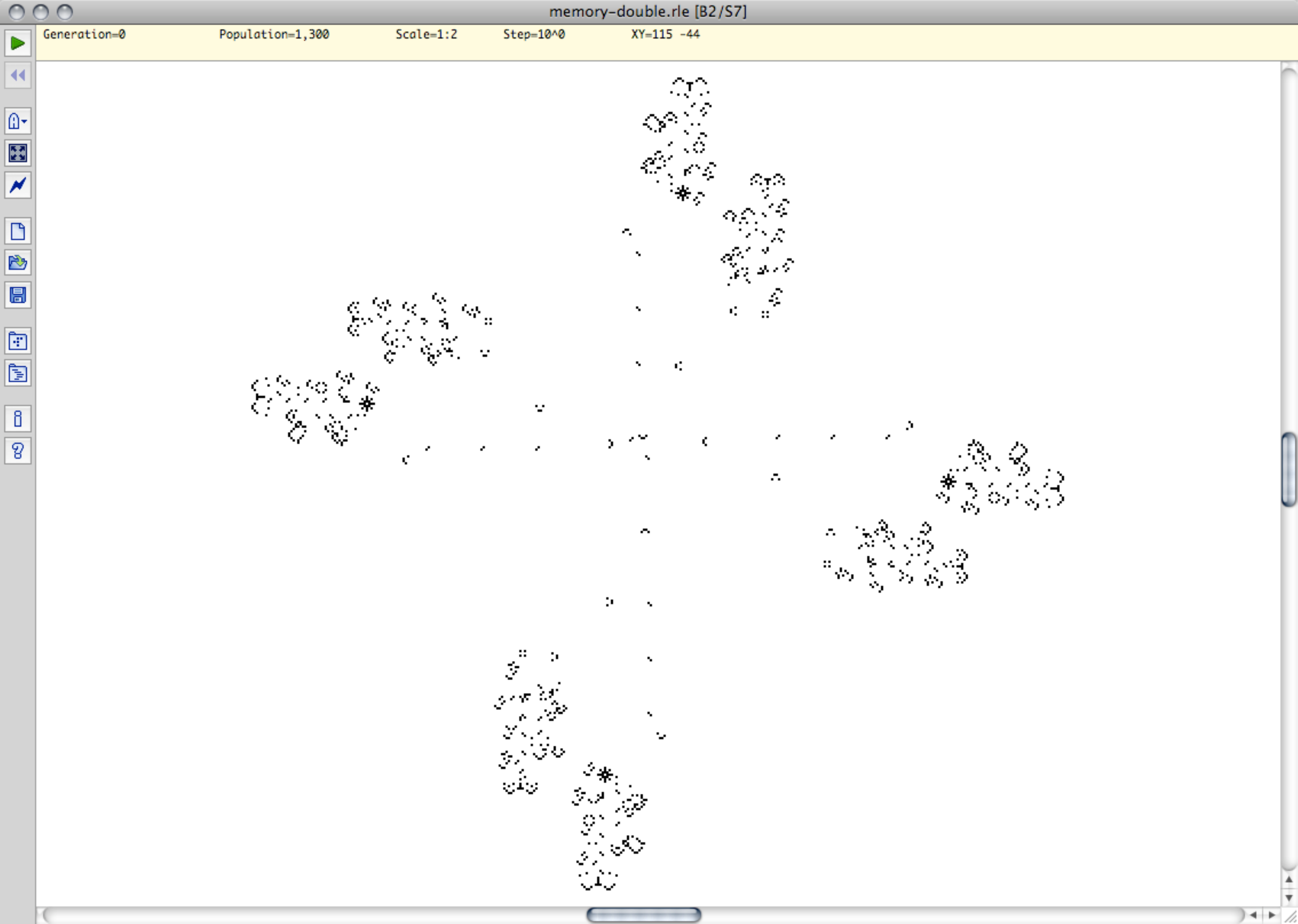}}} \hspace{0.8cm}
\subfigure[]{\scalebox{0.35}{\includegraphics{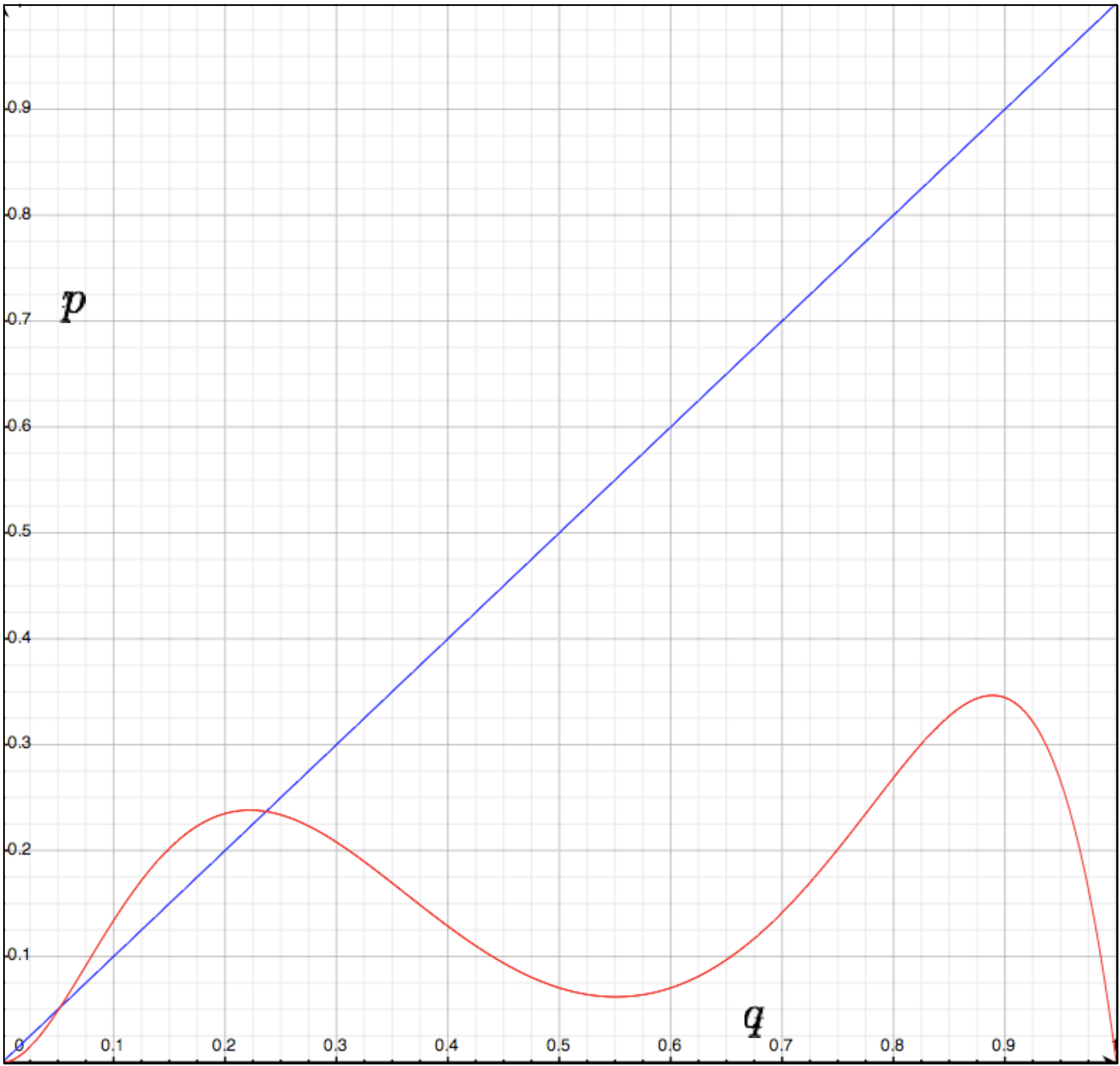}}}
\end{center}
\caption{(a) Memory device in the Diffusion Rule, (b) its mean field curve.}
\label{B2-S2345}
\end{figure}

 In this paper we present a construction in the Diffusion Rule that applies four bigger glider guns and four collisions to get a double memory device (that works equally well in the Seeds CA). Two small oscillators collide with four basic gliders to represent this primitive memory, and they are controlled by four identically composed glider guns rotated orthogonally, so that they travel forever in an infinite evolution space. Such a construction is shown in Fig.~\ref{B2-S2345}a. A number of collision analyses are in progress in a bid to obtain more complex computable devices.

 Thus Fig.~\ref{B2-S2345}b shows the mean field curve for the Diffusion Rule CA (or $B2/S7$) with polynomial:

$$
p_{t+1} = 4p^2_tq(2p^6_t+7q^6_t).
$$ 

The origin displays a stable fixed point (as in GoL) which guarantees the stable configuration in zero, while the unstable fixed point $p=0.05$ (significantly low compared to GoL) represents the densities where we find complex patterns emerging in the Diffusion Rule, as can be seen in Fig.~\ref{B2-S2345}b. 

The first maximum point $p=0.2381$ is very close to the second stable fixed point in $p=0.2363$ where the Diffusion Rule reaches its dominant density of live cells with a high level of activity--chaotic in this case, because the oscillation between the minimum $p=0.0618$ and the second maximum point $p=0.3464$ shows a different density in each generation, also oscillating to the second stable fixed point value.

\subsection{Life-like rule $B2/S2345$: Class III}
As we have seen, the computational universality of the GoL CA has already been demonstrated by various implementations. Among them are a functional register machine by Conway~\cite{kn:BCG82}, direct simulations of Turing machines by Paul Chapman~\cite{kn:Chap02} and Rendell \cite{kn:Ren02}, a complete set of logical functions by Jean-Philippe Rennard~\cite{kn:Renn02}, and recently, the design of a sophisticated universal constructor by Adam Goucher~\cite{kn:Gou09, kn:Gou10}. These implementations use principles of collision-based computing~\cite{kn:Ada02}, where information is transferred by mobile localizations (gliders) propagating in an architecture-less, or `free,' space. The theoretical results pertaining to GoL universality constitute just the first step in a long journey towards real-world implementation of collision-based computers as unconventional computing devices \cite{kn:Toff98, kn:Ada02}. Controllability of signals is the first obstacle to overcome. Despite their stunning elegance and complexity-wise efficiency of implementation, `free-space' computing circuits are difficult to fabricate from physical or chemical materials~\cite{kn:ACA05} because propagating localizations (solitons, breathers, kinks, wave-fragments, particles) are notoriously difficult to manipulate, maintain and navigate.

\begin{figure}
\begin{center}
\subfigure[]{\scalebox{0.47}{\includegraphics{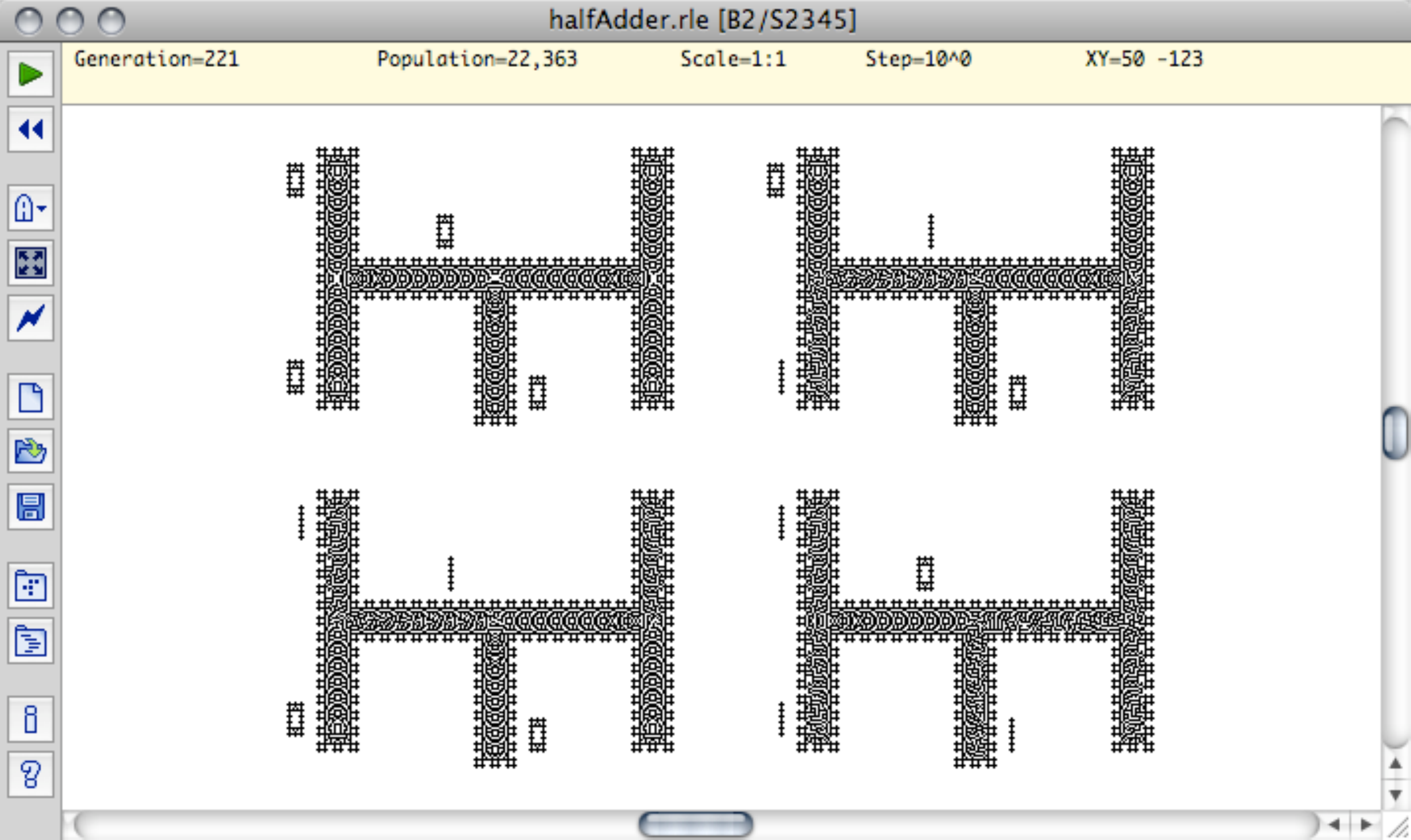}}} \hspace{0.8cm}
\subfigure[]{\scalebox{0.35}{\includegraphics{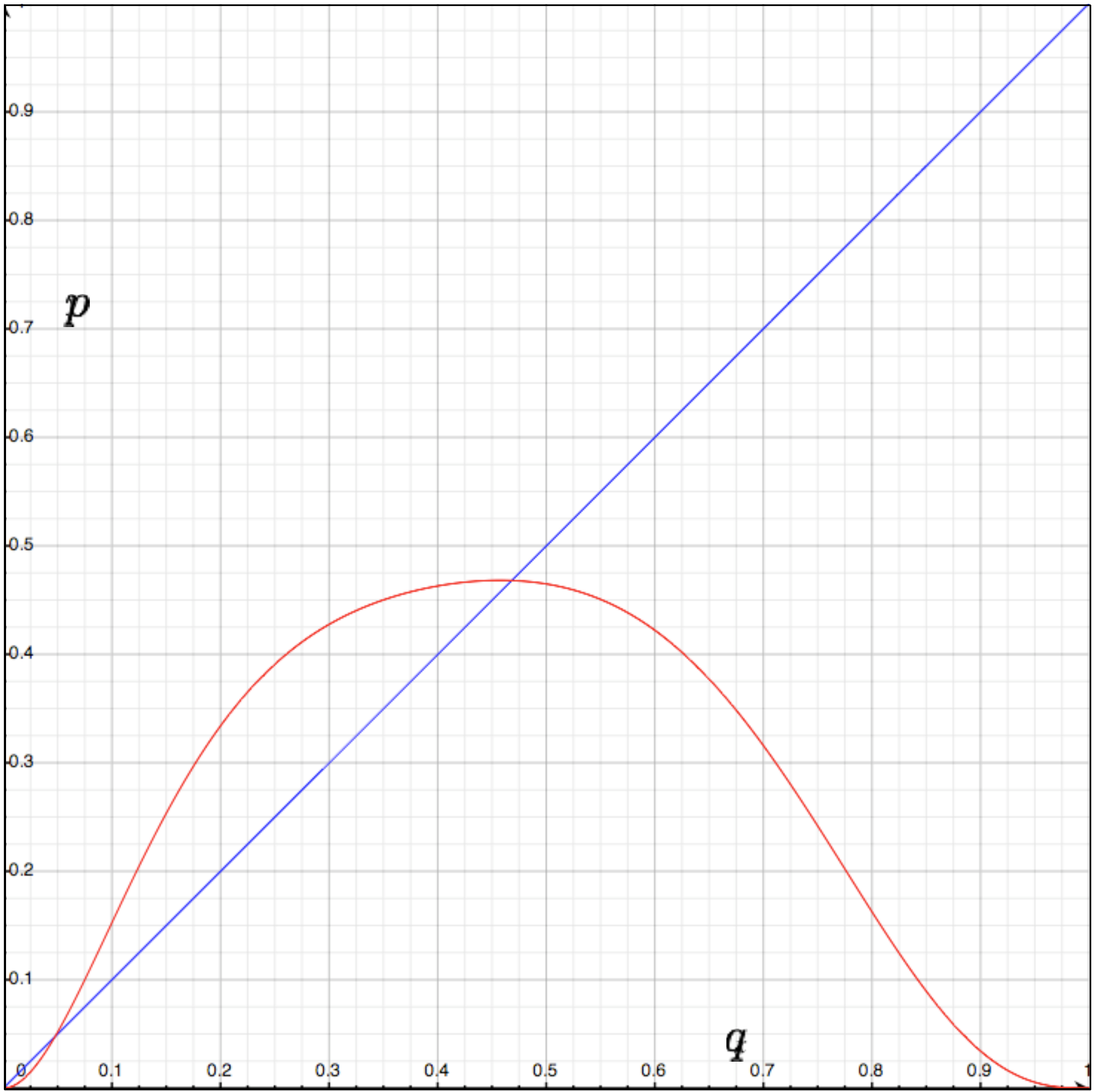}}}
\end{center}
\caption{Chaotic Life-like rule $B2/S2345$ (a) implementation of a half adder by competing patterns coded from glider reactions, (b) its mean field curve.}
\label{B2-S2345_HA_MF}
\end{figure}

In this section, we analyse the Life-like CA rule $B2/S2345$.\footnote{$B2/S2345$ is a CA within the domain of {\it Life-like rules} $dc22$ \url{http://uncomp.uwe.ac.uk/genaro/Life_dc22.html}}. This automaton is a discrete analog spatially extended chemical medium, combining properties of both sub-excitable and precipitating chemical media. From a random initial configuration, the $B2/S2345$ automaton exhibits chaotic behaviour, Class III. Configurations with low density of state `1' manifest the emergence of gliders and stationary localizations. This CA is able to support basic logic gates and elementary arithmetical circuits by simulating logical signals, with the propagation of gliders' growing geometrically restricted by stationary non-destructible localizations. Values of Boolean variables are encoded into two types of patterns --- symmetric ({\sc False}) and asymmetric ({\sc True}) patterns --- which compete for the `empty' space when propagating in the channels. Implementations of logic gates and binary adders are shown in \cite{kn:MAMM10, kn:MMAM10, kn:MAC08}. Thus Fig.~\ref{B2-S2345_HA_MF} depicts a binary half-adder implemented in $B2/S2345$. 

Figure~\ref{B2-S2345}b shows the mean field curve for $B2/S2345$ with polynomial:

$$
p_{t+1} = 7p^2_tq^3_t(4p_tq^3_t + 8p^2_tq^2_t + 10p^3_tq_t + 8p^4_t + 4q^4_t).
$$

The origin displays a stable fixed point (as in GoL) which guarantees the stable configuration in zero, while the unstable fixed point $p=0.0517$ (significantly low compared to GoL and close to that of the Diffusion Rule) represents the densities where we find complex patterns, as can be seen in Fig.~\ref{B2-S2345}b. The stable fixed point is very close to the maximum point. This is $p=0.4679$, almost a Gaussian curve.

\subsection{ECA Rule 110: Class IV}
\label{rule110}

The 1D binary CA rule numbered 110 in Wolfram's system of classification \cite{kn:Wolf83} has been the object of special attention due to the structures or gliders which have been observed in instances of its evolution from random initial conditions. The rule is assigned number 110 in Wolfram's enumeration because it represents the decimal base of the transition rule expanded in binary: 01110110. The transition function evaluates the neighbourhoods synchronously in order to calculate the new configuration transforming the neighbourhoods 001, 010, 011, 101 and 011 into state 1 and the neighbourhoods 000, 100 and 111 into state 0. It has been suggested that Rule 110 belongs to the exceptional Class IV of automata whose chaotic aspects are mixed with regular patterns. But in this case the background where the chaotic behaviour occurs is textured rather than quiescent, a tacit assumption in the original classification.\footnote{A repository of materials on ECA Rule 110 can be found at: \url{http://uncomp.uwe.ac.uk/genaro/Rule110.html}.} Rule 110 was granted its own appendix (Table 15) in \cite{kn:Wolf86}. It contains specimens of evolution including a list of thirteen gliders compiled by Lind and also presents the conjecture that the rule could be universal. 

The literature on the origins of Rule 110 includes a statistical study done by Wentian Li and Mats Nordahl in 1992 \cite{kn:LN92}. This paper studies the transitional role of Rule 110 and its relation to Class IV rules figuring between Wolfram's classes II and III. The study would seem to reflect an approach to equilibrium statistics via a power law rather than exponentially.

 Matthew Cook wrote an eight page introduction \cite{kn:Cook99} listing gliders from $A$ through $H$ and a glider gun.\footnote{An extended list of gliders in Rule 110 is provided in \url{http://uncomp.uwe.ac.uk/genaro/rule110/glidersRule110.html}.}. This list shows new gliders which do not appear on Lind's list, gliders with rare extensions, and a pair of gliders of complicated construction. Cook makes a comparison between Rule 110 and Life, finding some similarities in the behaviour of the two evolution rules and suggesting that Rule 110 may be called ``LeftLife.''

\begin{figure}
\centering
\includegraphics[width=1\textwidth]{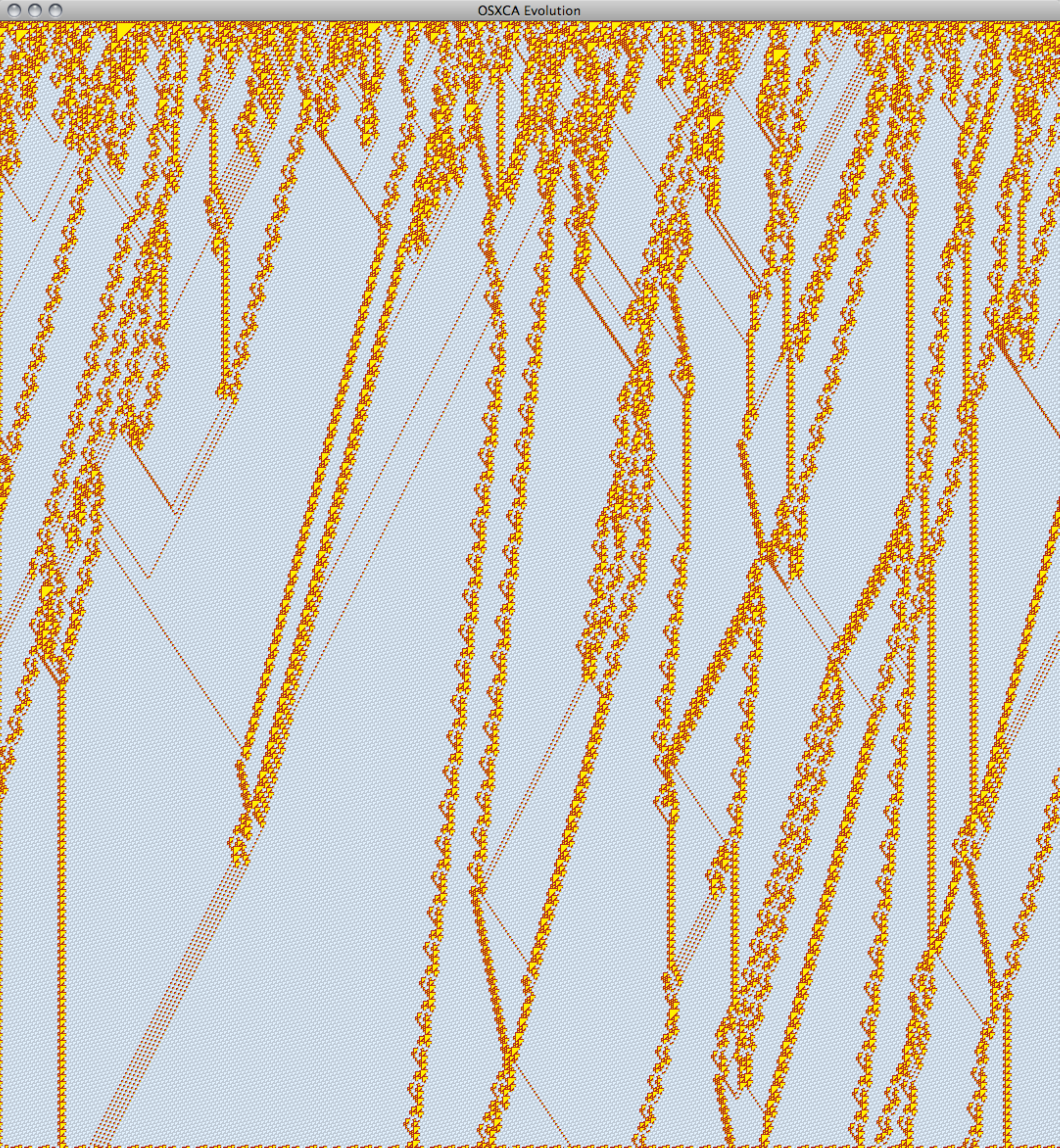}
\caption{Typical random evolution of ECA Rule 110. Initial density begins at 50\% per state in an evolution space of 1,244,300 cells. This is an initial condition of 1,082 cells evolving in 1,150 generations. A filter is selected for optimal clarity of gliders and collisions \cite{kn:MMS06, kn:MMS08}.}
\label{ECA110large_evol_rand}
\end{figure}

Looking at the rule itself, one notices a ubiquitous background texture which Cook calls ``ether,'' although it is just one of many regular stable lattices capable of being formed by the evolution rule, and can be obtained quickly using the de Bruijn diagrams \cite{kn:Mc99, kn:MMS06}. 

McIntosh raises the issue of the triangles of different sizes that cover the evolution space of Rule 110 \cite{kn:Mc00}. The appearance of these triangles suggests the analysis of the plane generated by the evolution of Rule 110 as a two dimensional shift of finite type. This suggestion is arrived at by observing that the basic entities in the lattices, the unit cells, induce the formation of upside-down isosceles right triangles of varying sizes. The significance of Rule 110 could lie in the fact that it is assembled from recognizably distinct tiles, and hence its evolution can be studied as a tiling problem, in the sense of Hao Wang \cite{kn:GS82}. It may even be possible to see fitting elements of one lattice into another as an instance of Emil L. Post's correspondence principle \cite{kn:Dav94}, which would establish the computational complexity of the evolution rule \cite{kn:Mc99}.

\begin{figure}
\centering
\includegraphics[width=1\textwidth]{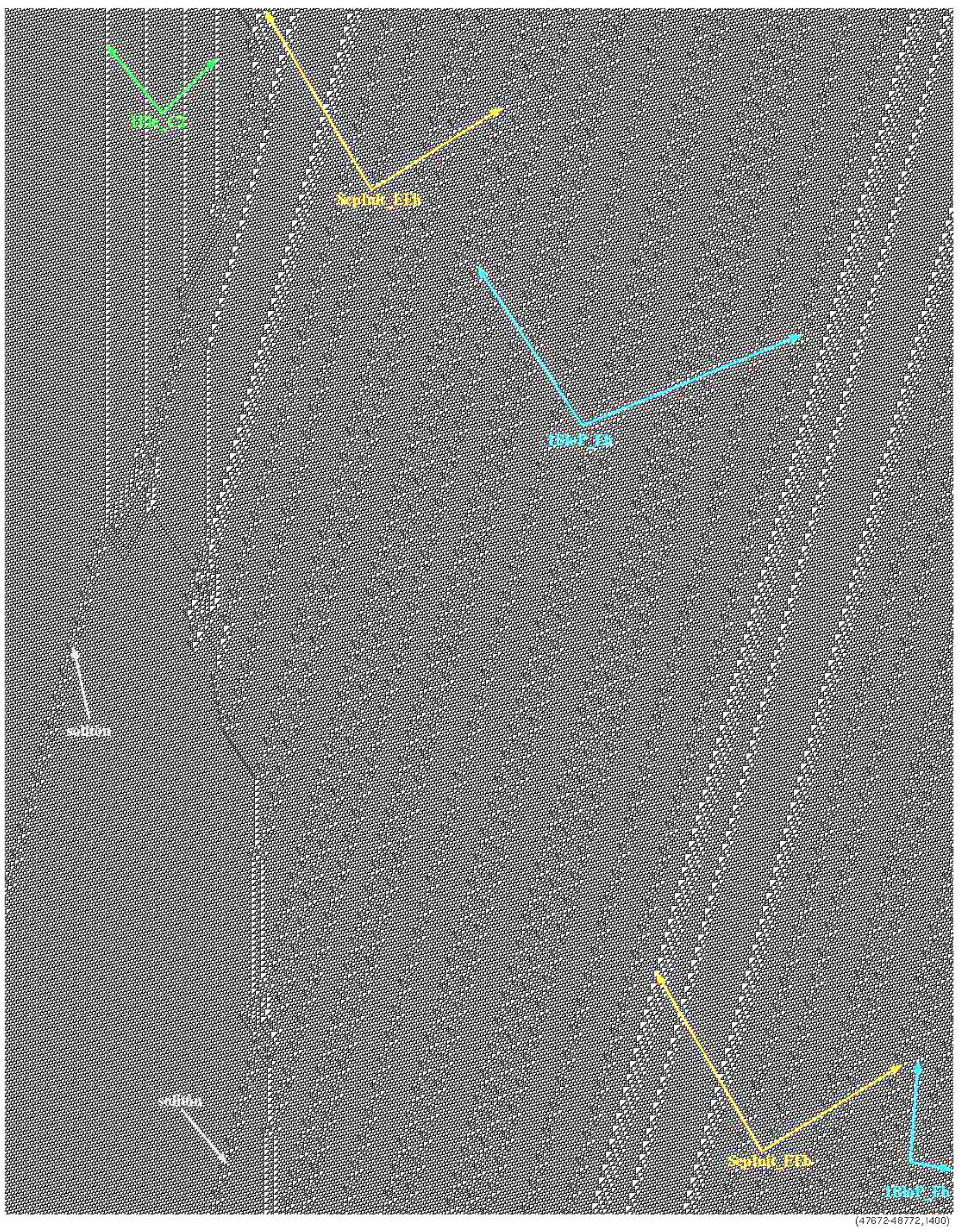}
\caption{First stage of a CTS working in Rule 110.}
\label{ctsCook-1}
\end{figure}

\begin{figure}
\centering
\includegraphics[width=1\textwidth]{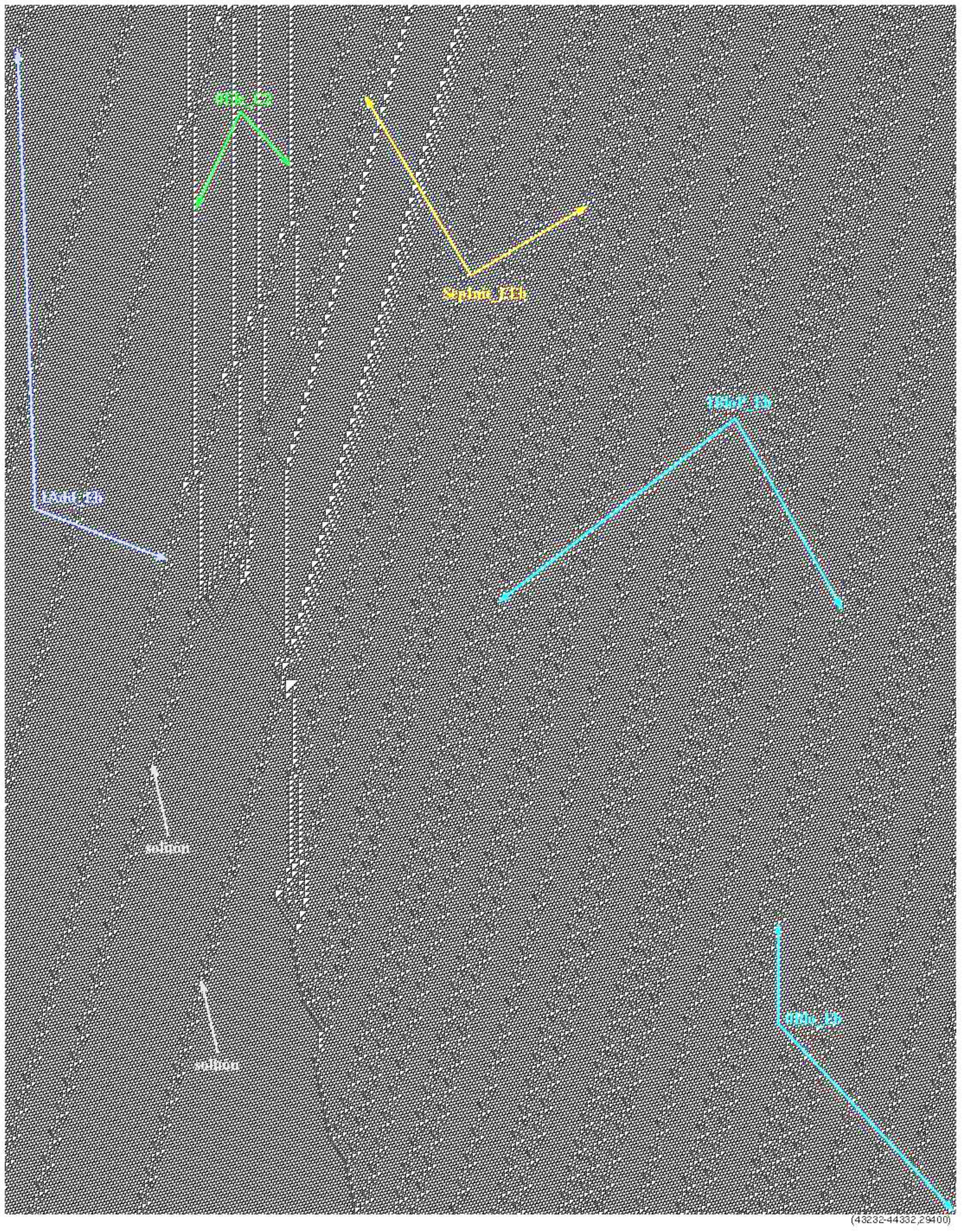}
\caption{Type, binary data (a 0 in this case), and deleting binary data in 29,400 generations.}
\label{ctsCook-8}
\end{figure}

\begin{figure}
\centering
\includegraphics[width=0.9\textwidth]{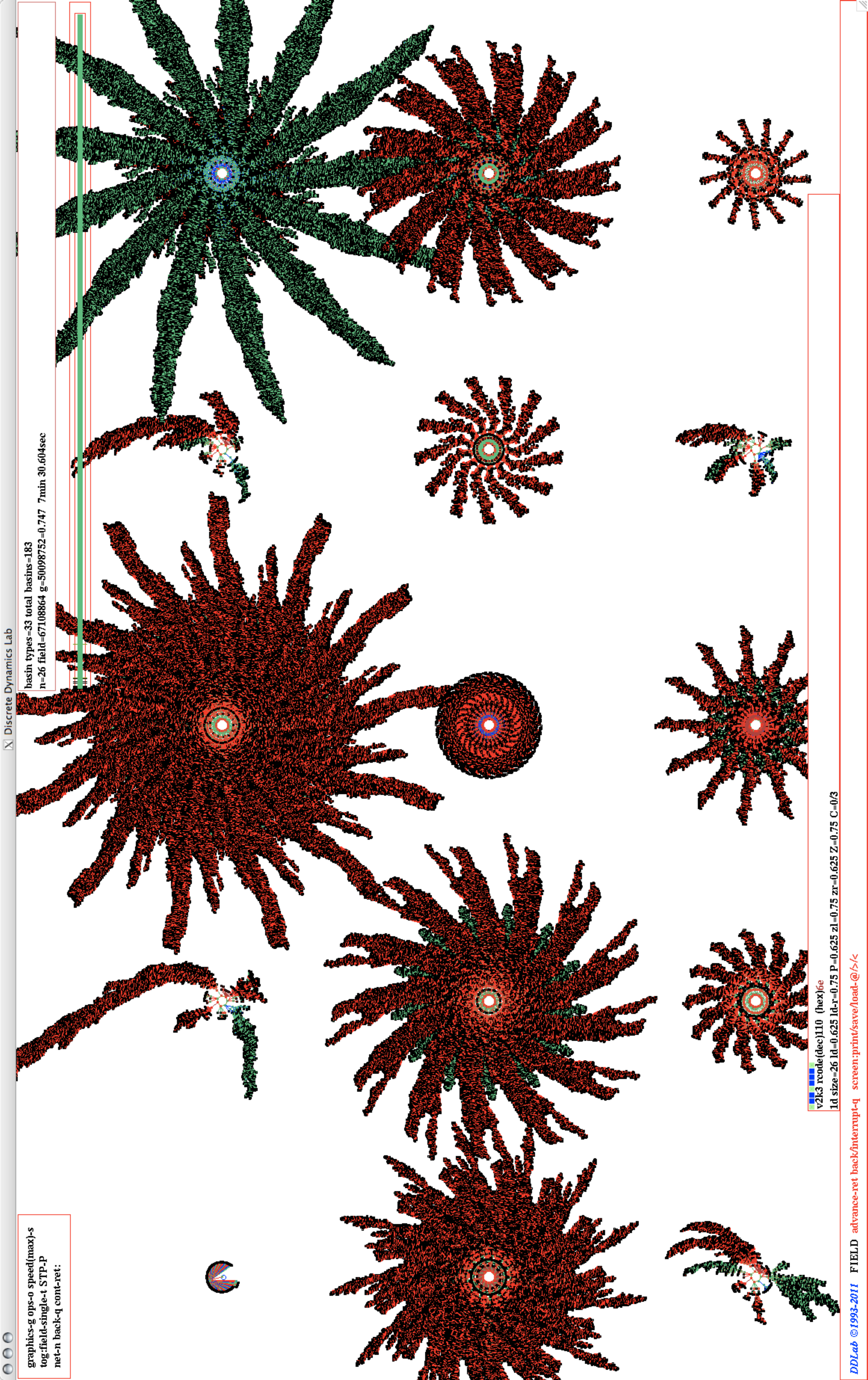}
\caption{Basin of attraction field for configurations on a ring of 26 cells in Rule 110.}
\label{basinECA110_26a}
\end{figure}

The most important result both in the study of Rule 110 and in CA theory over the last twenty years, is the demonstration that Rule 110 is universal \cite{kn:Cook04, kn:Wolf02, kn:Mc02, kn:Cook11, kn:MMS11}. 

In Fig.~\ref{ECA110large_evol_rand}, a typical random evolution of Rule 110 is displayed. Here we can see a diversity of gliders emerging and colliding for more than one thousand generations. The ether pattern is the periodic background where gliders travel and interact unperturbed. Consequently, as in GoL, each glider in Rule 110 can be obtained from a set of reactions among gliders, and {\it Rule 110 objects} can be constructed as well as Rule 110-based collisions (see \cite{kn:MMS07}).

 To show universality in Rule 110, a cyclic tag system (CTS) was designed to be useful in its particular environment with its characteristic restrictions: 1D, boundary conditions, package of gliders, and multiple collisions. CTS are new machines proposed by Cook in \cite{kn:Cook04} as tools for implementing computations in Rule 110. CTS are a variant of tag systems. Like the latter, they read a tape from the front and add characters at the end. Nevertheless there are some new characteristics and restrictions. Snapshots relating to their functionality are displayed in Fig.~\ref{ctsCook-1} and~\ref{ctsCook-8} \cite{kn:MMS11}.\footnote{A detailed description of this CTS working in Rule 110 can be found in \cite{kn:Wolf02, kn:Cook04, kn:Mc02, kn:Cook11, kn:MMS11}. Large high resolution snapshots of different stages of the machine are available on the Internet at \url{http://uncomp.uwe.ac.uk/genaro/rule110/ctsRule110.html}.}

\begin{figure}[th]
\centering
\includegraphics[width=0.6\textwidth]{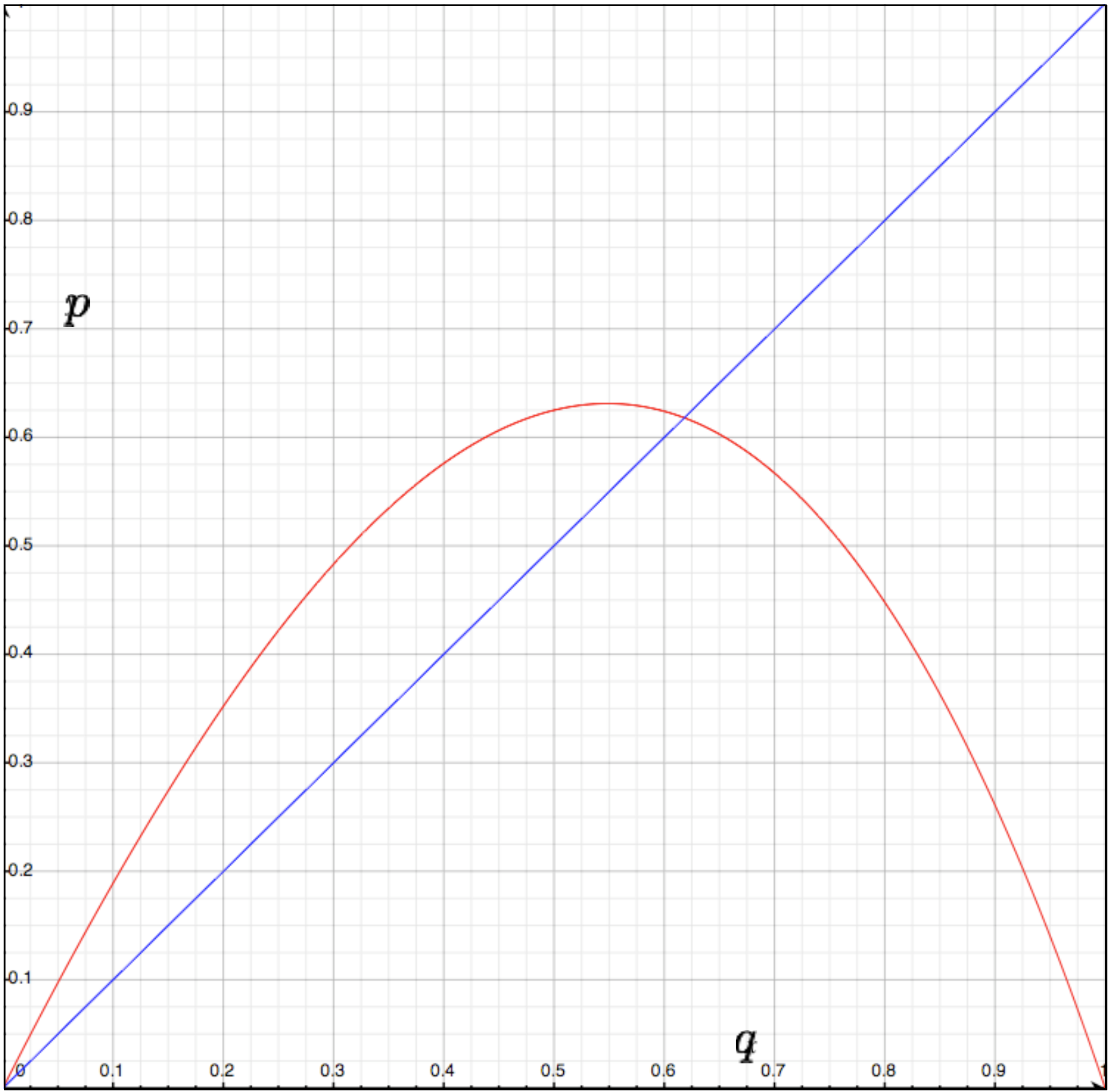}
\caption{Mean field curve for ECA Rule 110.}
\label{Rule110_MF}
\end{figure}

Thus Fig.~\ref{Rule110_MF}b shows the mean field curve for Rule 110 with polynomial:

$$
p_{t+1} = 2p_tq^{2}_t+3p^{2}_tq_t.
$$

The origin displays a stable fixed point (as in GoL) which guarantees the stable configuration in zero. The maximum point ($p=0.6311$) is close to the fixed stable point in $p=0.62$. In Rule 110 we cannot find unstable fixed points, and in any case the emergence of complex structures is ample and diverse.

 A basin (of attraction) field of a finite CA is the set of basins of attraction into which all possible states and trajectories will be organised by the local function $\varphi$. The topology of a single basin of attraction may be represented by a diagram, the {\it state transition graph}. Thus the set of graphs composing the field specifies the global behaviour of the system \cite{kn:WL92}.

 In Fig.~\ref{basinECA110_26a} we see the basin of attraction fields for a ring of 26 cells in Rule 110, the aptly named {\it cycle diagrams} \cite{kn:Mc09}. Wuensche determines how a basin of attraction field can classify CA into Wolfram's classes by means of attractors. For a CA Class IV we will see moderate transients, moderate-length periodic attractors, moderate n-degree, and very moderate leaf density \cite{kn:Wue99, kn:WL92}. In Fig.~\ref{basinECA110_26a} several cycles have symmetric ramifications. However, other cycles have non-symmetric ancestors with very long histories before they reach the root or attractor. Also the symmetric cycles have very long ramifications in comparison with chaotic rules, and the trees are not highly dense.

 As calculated in \cite{kn:Zen10}, rules such as Rule 110 and Rule 54 (also believed to be capable of universal computation, c.f. \ref{rule54}) had a large compression-based phase transition coefficient, as discussed in Section \ref{compressibility}, meaning that their ability to transfer information was well captured by the measure defined in \cite{kn:Zen10} (and, interestingly, perhaps strengthens the belief that Rule 54 is capable of Turing universality).

\subsection{ECA Rule 54: Class IV}
\label{rule54}

\begin{figure}
\centering
\includegraphics[width=1\textwidth]{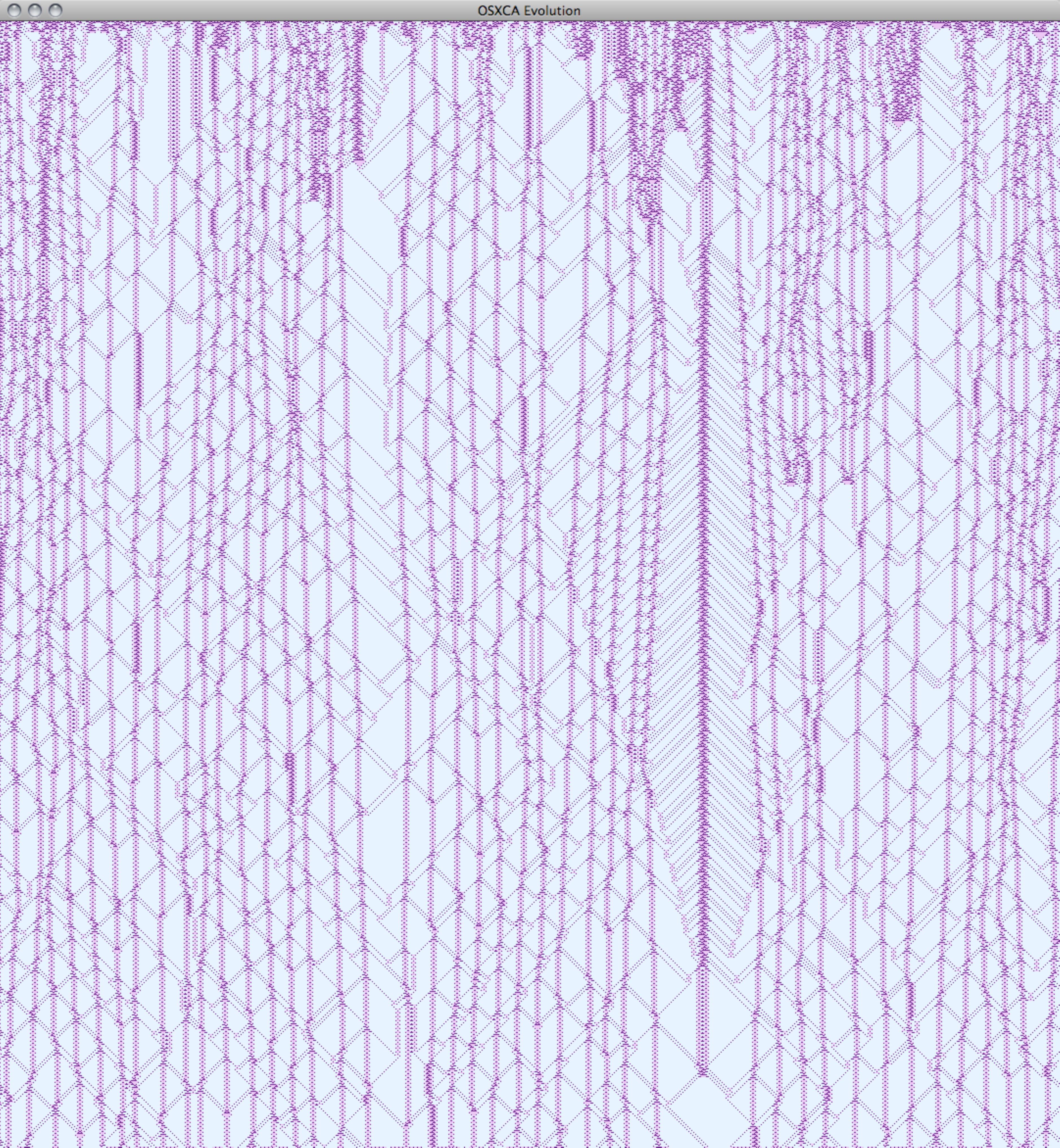}
\caption{Typical random evolution of ECA Rule 54. Initial density begins with 50\% per state in an evolution space of 1,244,300 cells. This is an initial condition of 1,082 cells evolving over 1,150 generations. A filter is selected for optimal clarity of gliders and collisions \cite{kn:MAM06, kn:MAC08}.}
\label{ECA54large_evol_rand}
\end{figure}

ECA Rule 54 is a two-state, three-neighbour cellular automaton in Wolfram's nomenclature, and is less complex than Rule 110. Nevertheless its dynamics are rich and complex.\footnote{A repository of materials on ECA Rule 54 can be found at: \url{http://uncomp.uwe.ac.uk/genaro/Rule54.html}.}  A Systematic and exhaustive analysis of glider behaviour and interactions, including a catalog of collisions, was provided in \cite{kn:MAM06}. Many of them promise computational elements for future designs. In one case a number of logic gates were derived from binary and triple collisions. In  \cite{kn:Wolf02}, Wolfram presents some functions produced by long series of periodic collisions in Rule 54 (page 697). However, no proof of the universality of Rule 54 has been offered yet. 

In their pioneering work, Boccara, Nasser, and Roger \cite{kn:BNR91} presented a preliminary list of gliders and discussed the existence of a glider gun. They also applied some statistical analysis to examine the stability of gliders. Later, Hanson and Crutchfield~\cite{kn:HC97} introduced the concept of ``computational mechanics'' -- applied finite state machine language representation-- in studying defect dynamics in 1D CA, and in deriving motion equations for filtered gliders. More studies were undertaken by Wolfram~\cite{kn:Wolf02}, who presented glider collisions with long periods of after-development and several filters for detecting gliders and defects, and Bruno Martin~\cite{kn:Mar00}, who designed an algebraic group of order four to represent collisions between basic gliders. By the way, David Hillman had calculated the same algebraic property before, but he did not publish it.\footnote{Personal communication.}

\begin{figure}[th]
\centering
\includegraphics[width=0.6\textwidth]{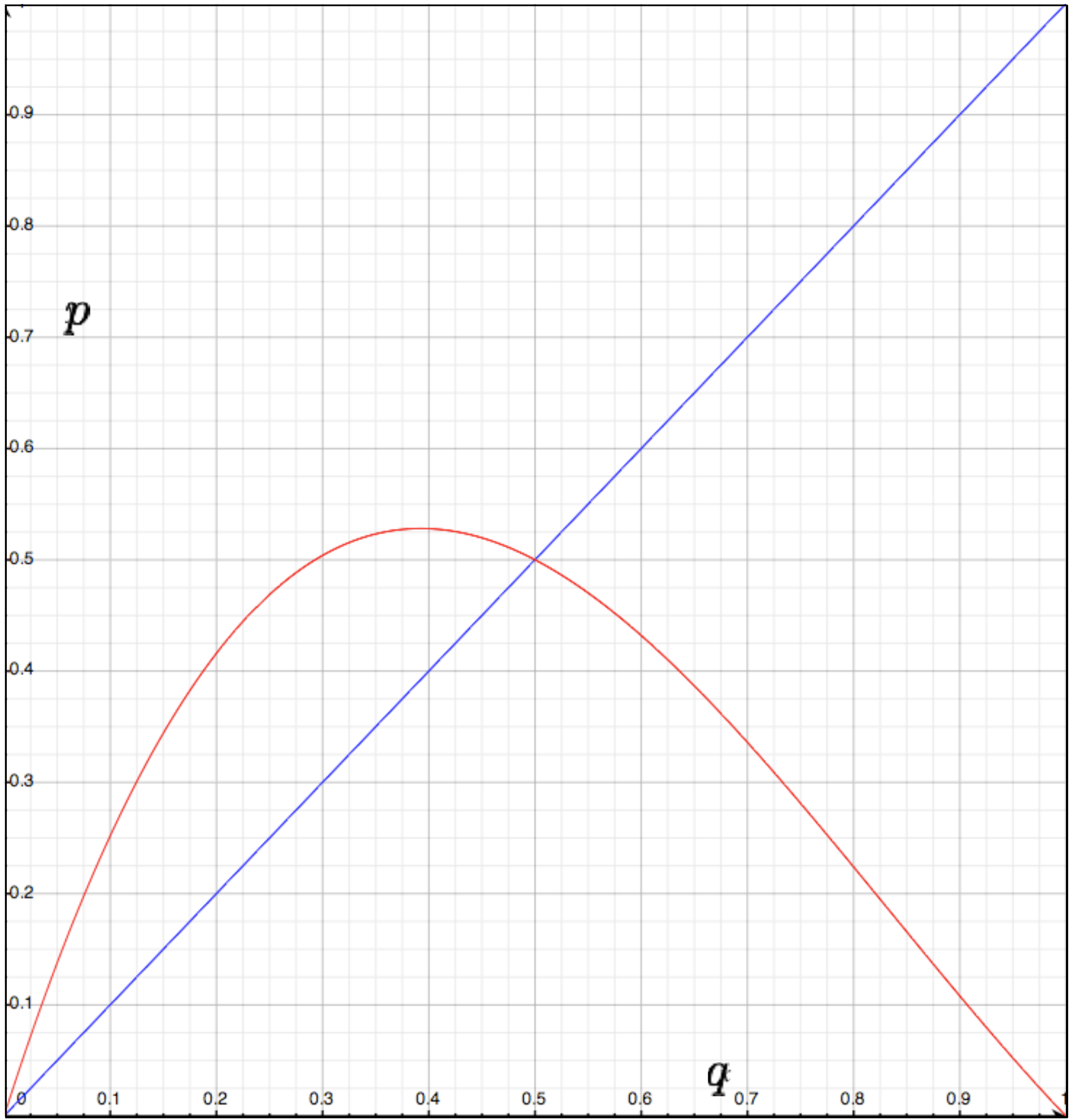}
\caption{Mean field curve for ECA Rule 54.}
\label{Rule54_MF}
\end{figure}

\begin{figure}
\centering
\includegraphics[width=0.9\textwidth]{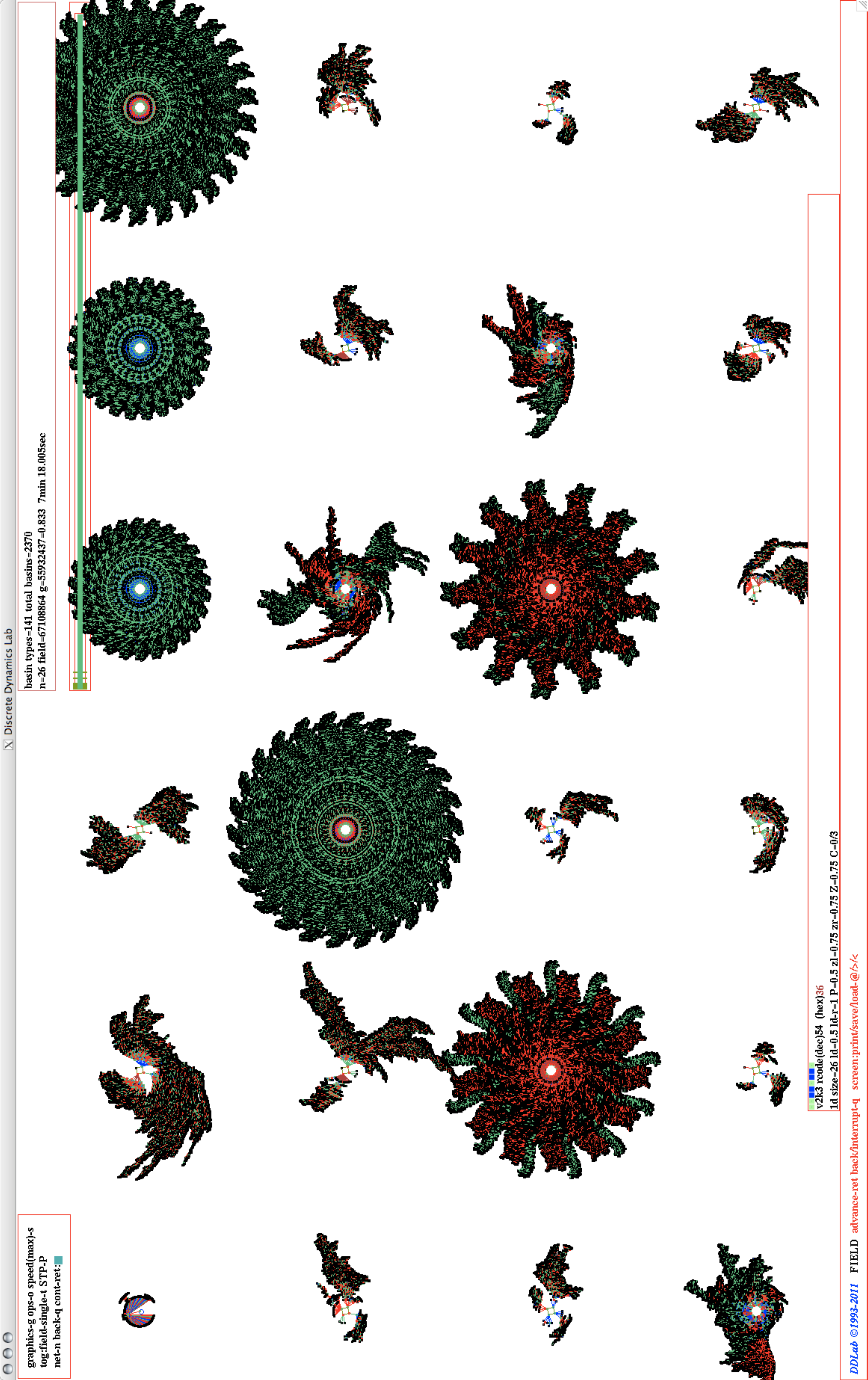}
\caption{Basin of attraction field for configurations on a ring of 26 cells in Rule 54.}
\label{basinECA54_26b}
\end{figure}

Figure~\ref{ECA54large_evol_rand} illustrates a typical random evolution in ECA Rule 54. The glider family comprises four basic gliders and one glider gun.\footnote{A full list of gliders in Rule 54 is available at \url{http://uncomp.uwe.ac.uk/genaro/rule54/glidersRule54.html}.} However, three new glider guns and other {\it Rule 54 object} based collisions were reported in \cite{kn:MAM06}, and an exotic double glider gun in \cite{kn:MAM08}. It is noteworthy that Rule 54 can yield glider guns from random initial conditions, which is very difficult in Rule 110. On the other hand, in Rule 110, sometimes the evolution reaches stability of complex structures between 100 to 500 generations, but in Rule 54 we have longer transients before achieving stability. Yet Rule 110 has a base of 12 gliders versus the four basic gliders in Rule 54. This instability is owed to the low probability of annihilation of gliders in Rule 54. Hence for binary collisions the probability is 10\% and for triple collisions it is 3.33\% \cite{kn:MAM06}.

Figure~\ref{Rule54_MF} shows the mean field curve for Rule 54 with polynomial:

$$
p_{t+1} = 3p_tq^2_t+p^2_tq_t.
$$

The origin displays a stable fixed point (as in GoL) which guarantees the stable configuration in zero. The maximum point ($p=0.5281$) is very close to the fixed stable point in $p=0.5$, although we cannot find unstable fixed points, as is the case with Rule 110 as well. 

On the other hand, in Fig.~\ref{basinECA54_26b} we can see the basin of attraction fields for a ring of 26 cells in Rule 54. Here we can see attractors with moderate transients, moderate-length periodic attractors, moderate n-degree, and very moderate leaf density. Also, we have some attractors with non-symmetric trees and branches, and we have other kinds of attractors with dense foliage strongly related to chaotic behaviour.

\section{Heat and programmability of Class III systems}
\label{heat}

In the Game of Life community there is an often used  \emph{heat} measure \footnote{See \url{http://www.argentum.freeserve.co.uk/life.htm} accessed in July 2012.}, defined as the average number of cells which change state in each generation (note the connections to Shannon's Entropy and the Mean Field Theory). For example, \emph{the heat} of a glider in GoL is known to be four, because two cells are born and two die in every generation, and that for a blinker is 4, because 2 cells are born and 2 die in every generation. In general, for a period $n$ oscillator with an $r$-cell rotor, the heat is at least $2r/n$, y and no more than $r(1-(n \mod 2)/n)$. 

Wolfram identified some of these issues in his enumeration of open problems in the research on CA \cite{wolframopen} (problems 1, 2 and 14), concerning the connections between the computational and statistical characteristics of cellular automata, measures of entropy and complexity and how to improve his classification using dynamic systems (which was one of the motivations of \cite{kn:Zen10}). Wolfram asks, for example, about the rate of information transmission of a CA in relation to its Lyapunov exponent (positive for Classes III and IV) and the computational power of these systems according to their classes.

The concept of \emph{heat} can clearly be associated with Wolfram's chaotic Class III, where CAs, e.g., rule 30, change state at a very high rate, (see Figures (c)~\ref{WolframClasses}), which is what keeps them from developing persistent structures such as are seen in Rule 110 (see Figure (d)~\ref{WolframClasses}, \ref{ctsCook-1} and \ref{ctsCook-8}). The presence of persistent structures in Wolfram's Rule 110 and Conway's Game of Life is what allows them to perform computation--implementing logic gates or transferring information over time by putting particles in the way of interacting with each other. So the question is whether CAs such as the ones belonging to Wolfram's Class III are too ``hot" to transfer information and are therefore, paradoxically, just like Class I systems--unable to perform computation. Alternatively, Class III may be able to perform computation, as has been suggested, but it may just turn out to be difficult to program such systems (if not designed to be a Class III system by using first a system from another class, somehow \emph{hiding} its computing capabilities), and this is what the compressibility approach discussed in Section \ref{compressibility} seems to be measuring for this class and which has been advanced in \cite{kn:ZenAISB} as a measure of \emph{programmability}.

\section{Final remarks}
\label{finalremarks}

Usually, chaotic rules are not considered candidates for computational universality. The Class III question we have formulated herein is the question of whether computation and Turing universality is possible in chaotic cellular automata. Universality results in simple programs capable of complicated behaviour have traditionally relied on localized structures (``particles") well separated by relatively uniform regions. This means that a measure like the entropy of the system tends to be well below its theoretical maximum. ÊThe open problem is therefore to prove computational universality in a simple program system for which an entropy measure on each time step remains near its maximum. Can a ``hot system" of this sort perform meaningful computation?

We have shown some cases where chaotic rules can support complex patterns, including logical universality. Exploring many CA rules, including the exceptionally chaotic Life-like rule {\it Dead without Life} \cite{kn:GM96}, one finds that there are several rules between chaos and complexity which are not included within the domain of complex behaviour. However, they present many elements equally likely to reach Turing computational universality. An important point made in this review is that it seems clearly to be the case that it is not only {\it complex CA}\footnote{A Complex Cellular Automata Repository with several interesting rules is available at \url{http://uncomp.uwe.ac.uk/genaro/otherRules.html}. We particularly recommend Tim Hutton's {\it Rule Table Repository} \url{http://code.google.com/p/ruletablerepository/}.} rules that are capable of computation, and that CA, even if simple or random-looking, may support Turing universality. Whether the encoding to make them actually compute turns out to be more difficult than taking advantage of the common interacting persistent structures in rules usually believed to belong to Wolfram's class IV is an open question. 

Previous results on universal CAs (developing signals, self-reproductions, gliders, collisions, tiles, leaders, etc.) prove that unconventional computing can be obtained depending on the nature of each complex system. For example, to prove universality in Rule 110 it was necessary to develop a new equivalent Turing machine to take advantage of limitations in 1D and the same dynamics in its evolution space, e.g., mobility of gliders and boundary properties. Hence, a CTS was devised, before this system was known as a circular machine \cite{kn:Arb69, kn:KR01, kn:Mor07, kn:MAS11}. This way, the nature of each system would determine the best environment in which to design a corresponding computer. This could be the basis of Wolfram's {\it Principle of Computational Equivalence} and it is also the inspiration behind the definition of \emph{programmability} measures for natural computation in \cite{kn:ZenAISB}. Wolfram's {\it Principle of Computational Equivalence} ultimately only distinguishes between two kinds of behaviours (despite Wolfram's own heuristic classification), namely those that are ``sophisticated" enough and reach \emph{Wolfram's threshold}, constituting a class of systems capable of computational universality, and those that fall below this threshold and are incapable of universal computation. And indeed, the compression-based classification in \cite{kn:Zen10} at first distinguishes only two classes. 

A number of approximations were developed or adapted to find complex CA. Perhaps the most successful technique was the one developed by Wuensche, with its $Z$ parameter \cite{kn:Wue99}. Some attempts were made by Mitchell {\it et. al} using genetic algorithms, although they had a particular interest in finding rules able to support complex patterns (gliders) with computational uses \cite{kn:DMC94, kn:WO08}. Unfortunately, these algorithms have strong limitations when it comes to searching in large rule spaces and very complex structures. And though the technique in \cite{kn:Zen10} has proven capable of identifying complex systems with great accuracy, it requires very large computational resources to extend the method to larger rule spaces if a thorough investigation is desired (though in conjunction with other techniques it may turn out to be feasible).

 As it has proven to be a very rich space, new kinds of CAs are proposed all the time. e.g., reversible CA \cite{kn:Kari96, kn:SHM08, kn:Mc91a}, partitioned CA \cite{kn:Wolf02}, hyperbolic CA \cite{kn:Mar07}, CA with non-trivial collective behaviour (self-organization) \cite{kn:CM92, kn:CGG08}, asynchronous CA \cite{kn:FM05}, biodiversity in CA \cite{kn:RAM13}, CA with memory \cite{kn:Alo09, kn:Alo11}, morphological diversity \cite{kn:AM10}, identification of CA \cite{kn:Ada94,kn:Zen10}, communication complexity \cite{kn:DRT04, kn:GMR11}, pattern recognition from CA \cite{kn:AN86}, to mention a few. 

Some other studies dedicated to designing or identifying universal CAs are \cite{kn:Hey98, kn:Ada01, kn:Ada02, kn:GM03, kn:MAS10}. Obtaining CA of Class IV from other rules has been studied via lattice analysis \cite{kn:Gunji10}, with memory \cite{kn:MAA10, kn:MAA12, kn:MAS10, kn:Alo03, kn:AM03, kn:Alo06}, asynchronous \cite{kn:Suzu94, kn:TV02, kn:CBM02, kn:FM05}, differential equations \cite{kn:Chua06}, partitioned \cite{kn:MH89, kn:Mor90, kn:IM00, kn:Mor07, kn:Mor08, kn:MMI99, kn:Marg98, kn:Marg03}, parity-filter CA \cite{kn:PST86, kn:Siw02, kn:JSS01}, number-conserving \cite{kn:MTI02} changing different neighbourhoods in CA \cite{kn:WN09}.

 CA as {\it super computer models} are developed extensively in \cite{kn:von66, kn:Codd68, kn:Ban71, kn:Marg84, kn:MTV86, kn:TM87, kn:Wolf88, kn:Sipp97, kn:Hey98, kn:Toff98, kn:FT01, kn:Ada01, kn:Ada02, kn:Ada02a, kn:ACA05, kn:Wors09, kn:Hutt10, kn:MAS11}.

\section*{Software used for simulations and plots in this paper}
\begin{myitemize}
\item Discrete Dynamics Lab (DDLab) \cite{kn:Wue11} \url{http://ddlab.org/}
\item Grapher \url{http://guides.macrumors.com/Grapher}
\item Golly \url{http://golly.sourceforge.net/}
\item OSXCA system \url{http://uncomp.uwe.ac.uk/genaro/OSXCASystems.html}
\item Wolfram \emph{Mathematica} 8 \url{http://www.wolfram.com/}
\end{myitemize}

\section*{Acknowledgement}
G. J. Mart{\'i}nez wants to thank support given by EPSRC grant EP/F054343/1, J. C. Seck-Tuoh-Mora wants to thank support provided by CONACYT project CB-2007-83554 and H. Zenil wants to thank support by the FQXi under grant number FQXi-MGA-1212.


\end{document}